\documentclass[twocolumn]{aastex62}

\usepackage{apjfonts}
\usepackage[T1]{fontenc}
\usepackage{color}
\usepackage{amsmath,amstext}
\usepackage{hyperref}
\usepackage{natbib}
\usepackage{graphicx}
\usepackage{gensymb}

\begin{document}

\title{\sc GRB~191019A: a short gamma-ray burst in disguise from the disk of an active galactic nucleus} 

\author{Davide Lazzati}
\affil{ Department of Physics, Oregon State University, 301
  Weniger Hall, Corvallis, OR 97331, USA}

\author{Rosalba Perna} 
\affil{Department of Physics and Astronomy, Stony Brook
  University, Stony Brook, NY 11794-3800, USA}
\affil{Center for Computational Astrophysics, Flatiron Institute, New York, NY 10010, USA}

\author{Benjamin P. Gompertz}
\affil{School of Physics and Astronomy \& Institute for Gravitational Wave Astronomy, University of Birmingham, Birmingham, B15 2TT, UK}

\author{Andrew J. Levan}
\affil{Department of Astrophysics/IMAPP, Radboud University Nĳmegen, P.O. Box 9010, 6500 GL Nĳmegen, The Netherlands}
\affil{Department of Physics, University of Warwick, Coventry, CV4 7AL, UK}

\begin{abstract}

Long and short gamma-ray bursts (GRBs), canonically separated at around 2~seconds duration, 
are associated with different progenitors: the collapse of a massive star and the merger of two
compact objects, respectively. GRB~191019A was a long GRB ($T_{90}\sim 64$~s). Despite the relatively small redshift $z=0.248$ and HST followup observations,  an accompanying supernova was not detected. In addition, the host galaxy did not have significant star formation activity. Here we propose that GRB~191019A was produced by a binary compact merger, whose prompt emission was stretched in time by the interaction with a dense external medium. This would be expected if the burst progenitor was located 
in the disk of an active galactic nucleus, as supported by the 
burst localization close to the center of its host galaxy. We
show that the light curve of GRB~191019A can be well modeled 
by a burst of intrinsic duration $t_{\rm eng}=1.1$~s and of energy $E_{\rm iso}= 10^{51}$~erg seen moderately off-axis, exploding in a medium
of density $\sim 10^7-10^8$~cm$^{-3}$. 
The double-peaked light curve carries the telltale features
predicted for GRBs in high-density media, where the first peak is produced by
the photosphere, and the second by the overlap of reverse shocks that take place before the internal shocks could happen. This would make GRB~191019A the first confirmed stellar explosion from within an accretion disk, with important implications for the formation and evolution of stars in accretion flows and for gravitational waves source populations.

\end{abstract}
\section{Introduction}
\label{intro}

Long and short gamma-ray bursts (GRBs) have been traditionally
associated with galactic environments, where ambient densities
are generally small or moderate (few to hundreds of protons per cm$^{-3}$).
However, the possibility that a fraction of them may originate in
the disks of active galactic nuclei (AGNs) has been gaining some interest, especially in light of the fact that
AGN disks have been shown to be a
promising channel to explain some unexpected findings of the
LIGO/Virgo data, such as BHs both in the low
\citep{Abbott2020low,Yang2020,Tagawa2020} and in the high mass
gap \citep{Abbott2020high}, as well as an asymmetry in the inferred BH spin
distribution \citep{Callister2021,McKernan2021,Wang2021a}.

The presence of neutron stars (NSs) and black holes (BHs) embedded in
AGN disks is not surprising since they are the remnants of massive
stars.  Stars  are  expected to be
present in these disks as a result of two processes: capture from the
nuclear star cluster surrounding the AGN
(e.g. \citealt{Artymowicz1993}), and in-situ formation from
gravitational instabilities in the outer disk
(e.g. \citealt{Goodman2003,Dittmann2020}). Once formed or captured,
stars in AGN disks follow different evolutionary paths compared to
their galactic counterparts. The very large densities and strong torques
of the AGN disk environments cause stars embedded in them to both grow
to large masses \citep{Cantiello2021,Dittmann2021} as well as to acquire
angular speed \citep{Jermyn2021}, which makes them ideal
candidates as progenitors of long GRBs.

Additionally, and especially important for our investigation,  short GRBs are expected to explode in AGN disks. Short GRBs are
known to be the result of a merger of two NSs \citep{AbbottGW17}, and
potentially also of a NS-BH merger for non-extreme mass ratios.
NSs can be formed in AGN disks either in situ from the direct collapse
of stars, or as a result of capture from the nuclear star cluster
\citep{Tagawa2020,Perna2021b}.  Binary formation via dynamical
interactions is then facilitated by gas drag \citep{Tagawa2020} as
well as by compact object clustering in migration traps
(e.g. \citealt{Bellovary2016,McKernan2020}).

The medium of an AGN disk, due to its high density, can
however significantly change the appearance of a GRB.
Depending on the disk size and location within the disk,
GRBs can be choked \citep{Zhu2021a} or appear diffused
\citep{Perna2021,Wang2022}. In some cases they would also 
produce luminous neutrino bursts \citep{Zhu2021nu}. Time-dependent photoiozation of
the intervening material up to the photosphere can further alter early-time emission \citep{Ray2023}. In addition, the high density of the medium can dramatically change the
{\em intrinsic} spectra and light curves.  \citet{Lazzati2022} showed that in
high-density environments GRBs
are likely characterized by a single, long
emission episode that is due to the superposition of individual
pulses, with a characteristic hard-to-soft evolution irrespective of
the light-curve luminosity. Short duration GRBs can become long GRBs.
Additionally, a distinctive feature would be the lack of an associated supernova component and a position coincident with the center of the host galaxy.

A  burst with the key features of a GRB with a compact merger engine from an AGN disk has been recently identified \citep{Levan2023} in
GRB~191019A, as described in the next section. 
Here, using our formalism for the computation of the light curve of GRBs
in very dense media \citep{Lazzati2022}, we model the light curve of GRB~191019A
and derive the properties of its engine, as well as the density of the medium,
which we find consistent with that of an AGN disk.

The paper is organized as follows: Sec.~2 describes the observations
of GRB~191019A; our numerical
methods for the light curve modeling are described in Sec.~3. 
The simulations results are presented
in Sec.~4, and we summarize and put our results in context in Sec.~5.

\section{GRB~191019A}
\label{sec:grb}

GRB~191019A was detected by the \textit{Neil Gehrels Swift Observatory} on 19 October 2019. Its duration of $T_{90}=64.4\pm 4.5$~s indicated its belonging to the class of long GRBs. The light curve was characterized by a fast rise and an exponential decay (a.k.a.\@ FRED), with some evidence of hard-to-soft spectral evolution and overlaid variability. Optical observations with the Nordic Optical Telescope revealed a faint optical transient whose location was pinpointed  within $\sim 100$~pc of the nucleus of its host galaxy, whose redshift was measured to be $z=0.248$ \citep{Levan2023}. 

Despite the relatively low redshift of the burst and deep follow up observations with the \textit{Hubble Space Telescope}, no supernova counterpart was identified, making  its classification as a long GRB questionable.
Additionally, the lack of evidence for star formation in the host galaxy cast further doubt on its origin from the collapse of a massive star.
\cite{Levan2023} suggested that GRB~191019A is rather the result of the merger of two compact objects, involving white dwarfs, NSs or BHs, and that dynamical interactions in the dense cluster surrounding the central supermassive black hole of the host galaxy are responsible for forming the binary which eventually merged.

Our interpretation is broadly consistent with theirs in that we consider that GRB~191019A is rather the result of a binary merger than a collapsar, and that it originated in an environment prone to binary formation via dynamical interactions.
However, we make our suggestion more specific by proposing that GRB~191019A is a rather typical short GRB emerging from the disk of an AGN, and, as shown in the following, we support it by  demonstrating that its light curve and spectral evolution can be modeled as that of a short GRB exploding in a very high density medium.


\section{Methods}
\label{sec:methods}

We model the light curve of the prompt emission of GRB~191019A based on the model developed in~\citet{Lazzati2022}. To improve on their original setup, we also consider the emission from the fireball's photosphere. In this section we describe in detail 
the model for the photospheric component and briefly summarize the \cite{Lazzati2022}
model for the synchrotron shock component.

Photospheric emission in GRB fireballs has been extensively studied, both analytically
\citep{Peer2005,Giannios2006,Peer2006,Giannios2007} and numerically \citep{Chhotray2015,Ito2015,Lazzati2016,Parsotan2018,Ito2021}. Let us consider a fireball that is launched through a nozzle at a radius $r_0$ with bulk Lorentz factor $\Gamma_0=1$ and isotropic luminosity $L_{\rm{iso}}$. Its nozzle temperature is
\begin{equation}
    T_0=\left( \frac{L_{\rm{iso}}}{4\pi r_0^2 a c}\right)^\frac14,
    \label{eq:t0}
\end{equation}
where $a=7.56\times10^{-15}$~erg~cm$^{-3}$~K$^{-4}$ is the radiation density constant. The temperature decreases with distance due to the fireball's acceleration. At the saturation radius $r_{\rm{sat}}$, where the fireball ends its acceleration, it has reached a value
\begin{equation}
    T_{\rm{sat}} = \frac{T_0}{\eta},
\end{equation}
where $\eta=L_{\rm iso}/(\dot{m}c^2)$ is the fireball's asymptotic Lorentz factor. Beyond saturation, the temperature drops adiabatically
until the photospheric radius is reached, at which point the advected radiation is released to form the photospheric component of the prompt light curve. The temperature at the photospheric radius therefore reads (e.g., \citealt{Piran2004}):
\begin{equation}
    T_{\rm{ph}}=T_{\rm{sat}} \left(\frac{r_{\rm{ph}}}{r_{\rm{sat}}}\right)^{-\frac23}
    =
    \frac{T_0}{\eta} \left(\frac{r_{\rm{ph}}}{r_{\rm{sat}}}\right)^{-\frac23}.
    \label{eq:tph}
\end{equation}
The above equation gives the fireball temperature at the stage in which the radiation is released. Because of the blueshift due to the bulk motion towards the observer, the observed color temperature of the radiation is
\begin{equation}
    T_{\rm{obs}}=\eta T_{\rm{ph}} = 
    T_0\left(\frac{r_{\rm{ph}}}{r_{\rm{sat}}}\right)^{-\frac23},
\end{equation}
which depends only on the fireball's initial conditions through $T_0$ and on the two 
characteristic radii $r_{\rm{sat}}$ and $r_{\rm{ph}}$. For the latter, we use Eq. (6) from \cite{Lazzati2020}, which includes both thin and thick fireballs and incorporates the possibility of 
an electron fraction $Y_e\le1$. The smaller the electron fraction, the lower the Thomson thickness of the fireball and the sooner the photospheric component is released, causing an earlier and hotter photospheric component. 

To calculate $r_{\rm{sat}}$ we follow the standard fireball model \citep{Piran2004}, but we introduce a parameter $\alpha_{\rm{acc}}$ that controls the efficiency of the acceleration. In a standard fireball, evolving in vacuum, the fireball is spherical or conical (if beamed) and the acceleration is linear with distance: $\Gamma(r)=r/r_0$. Theoretical considerations \citep{Matzner2003,Bromberg2007,Lazzati2019,Gottlieb2022} and numerical simulations for both collapsars and binary neutron star mergers \citep{MacFadyen1999,Aloy2000,Morsony2007,Murguia2014,Nagakura2014,Gottlieb2021} have however shown that a jet expanding in an external medium is shocked and hydrodynamically collimated. This non-conical evolution delays acceleration, causing the saturation radius to happen at a greater distance, therefore reducing the length scale over which adiabatic cooling takes place, and eventually causing a brighter and hotter photosphere. We parameterize this deviation from conical evolution through a parameter $\alpha_{\rm{acc}}$ which is intrinsically defined through:
\begin{equation}
    \Gamma(r)=\left(\frac{r}{r_0}\right)^{\alpha_{\rm{acc}}},
\end{equation}
which modifies the equation for the saturation radius as:
\begin{equation}
    r_{\rm{sat}}=r_0\,\eta^\frac{1}{\alpha_{\rm{acc}}}.
    \label{eq:alpha}
\end{equation}
Putting it all together we arrive at the equations for the observed energetics of the photospheric emission, its observed peak frequency, and the observed duration of the pulse. The comoving energetics is given by the black body radiation density times the fireball volume, which is boosted by $\eta^2$ to calculate the observed isotropic equivalent energy:
\begin{equation}
    E_{\rm{ph,iso}}=4\pi r_{\rm{ph}}^2 c t_{\rm{eng}} a T_{\rm{ph}}^4 \eta^2,
\end{equation}
where $t_{\rm{eng}}$ is the time the central engine is active. Using Equations~\ref{eq:t0},~\ref{eq:tph}, and~\ref{eq:alpha}, we obtain:
\begin{equation}
    E_{\rm{ph,iso}}=L_{\rm{iso}}\left(\frac{r_0}{r_{\rm{ph}}}\right)^\frac23 \eta^\frac{8-6\alpha_{\rm{acc}}}{3\alpha_{\rm{acc}}} t_{\rm{eng}}\,.
\end{equation}
Analogously, the peak photon energy in keV is obtained as:
\begin{eqnarray}
    h\nu_{\rm{pk}} &=& \frac{2.8}{1.6\times10^{-9}} k_B T_{\rm{ph}} \eta = \nonumber \\ 
    &=&1.75\times10^9 k_B \left(\frac{L}{4\pi a c}\right)^\frac14 \frac{r_0^\frac{5}{12}}{r_{\rm{ph}}^\frac23} \eta^\frac{2}{3\alpha_{\rm{acc}}},
\end{eqnarray}
where $k_B$ is the Boltzmann constant. Finally, the duration of the photospheric pulse is set either by the engine duration $t_{\rm{eng}}$ or by the curvature timescale, whichever is longer:
\begin{equation}
    \Delta t_{\rm{ph}} = \max\left(t_{\rm{eng}},\frac{r_{\rm{ph}}}{c\eta^2}\right).
\end{equation}

\begin{figure}
\includegraphics[width=\columnwidth]{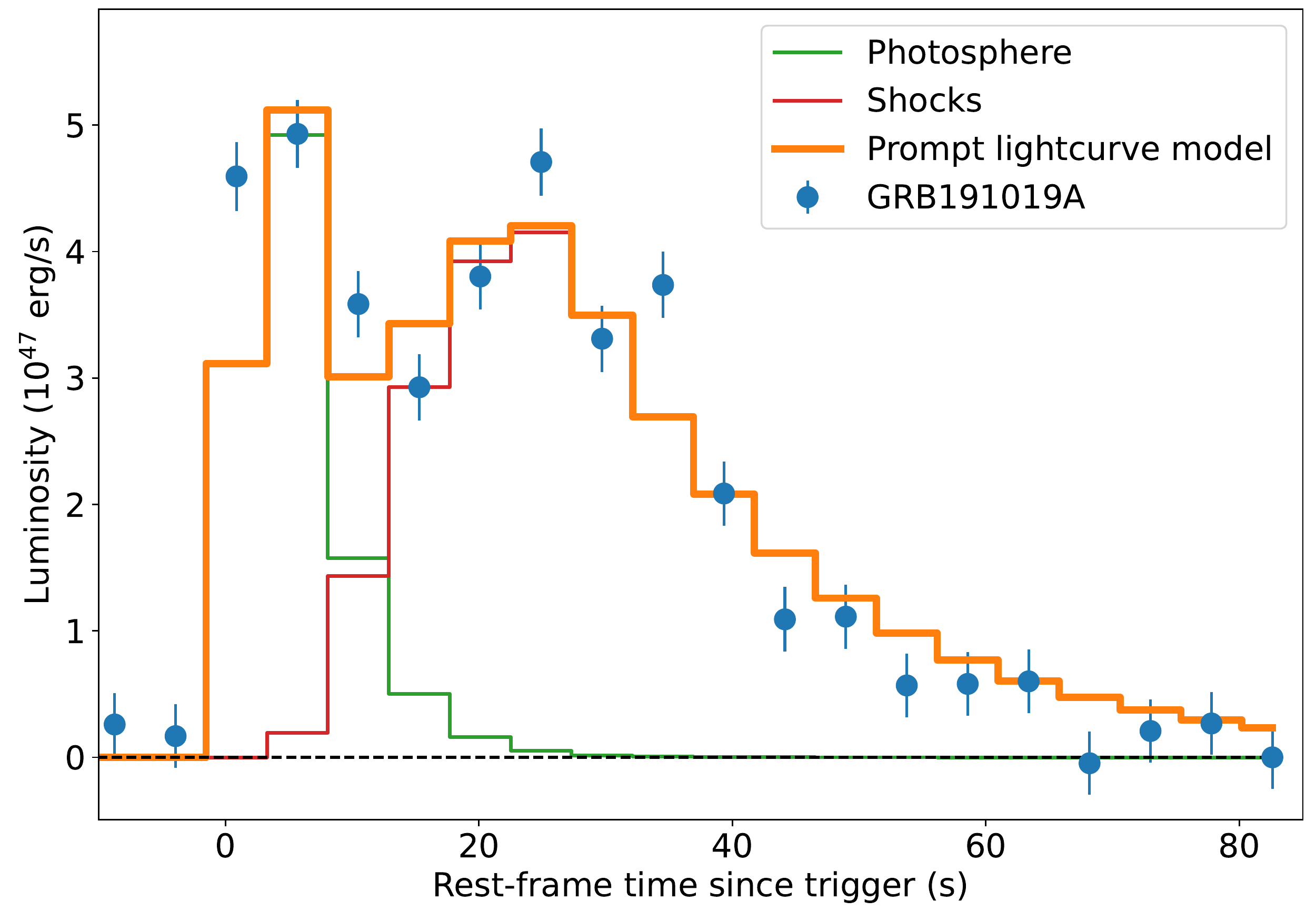}
\caption{Best fit model for the prompt light curve of GRB~191019A. \textit{Neil Gehrels Swift Observatory} data (blue symbols) are adapted from \cite{Levan2023}. The thick orange line shows the best fit model for the overall prompt light curve. The thin green line marks the 
photospheric component, while the thin red line displays the contribution of the strong reverse shocks driven into the fireball by the dense external medium.}
\label{fig:bestfit}
\end{figure}

Following the photospheric pulse is the train of pulses caused by the repeated shocking of the external medium caused by the incoming fireball shells. The unique feature of the model is that, due to the high external medium density, the external shocks take place very close to the central engine, at a distance that is smaller than the one at which internal shocks are expected. \cite{Lazzati2022} show that this causes a train of pulses whose duration is significantly longer than their spacing, creating a broad pulse
with hard to soft spectral evolution. In the following we model the prompt light curve of GRB~191019A as the superposition of the photospheric pulse discussed above and the train of shocks discussed in \cite{Lazzati2022}.

The particular setup chosen for the modeling has a central engine with $t_{\rm{eng}}$ fixed at 1.1~s. The fireball is made of six 0.1~s pulses with identical properties separated by dead times of 
0.1~s duration. Each pulse carries 1/6 of the overall energy (which is a fit parameter) and has an
asymptotic Lorentz factor $\eta$ (also a fit parameter). The nozzle radius $r_0$ is fixed at $r_0=10^{8}$~cm. The external medium is assumed to be uniform with density $n_{\rm{ext}}$ (a fit parameter). The microphysical parameters for computing the radiation efficiency are held fixed at $\varepsilon_e=0.2$ and $\varepsilon_B=0.01$. The electron fraction $Y_e$ and the acceleration efficiency $\alpha_{\rm{acc}}$ are the two final fit parameters.

\section{Modeling Results}
\label{sec:results}

\begin{figure*}
\includegraphics[width=\textwidth]{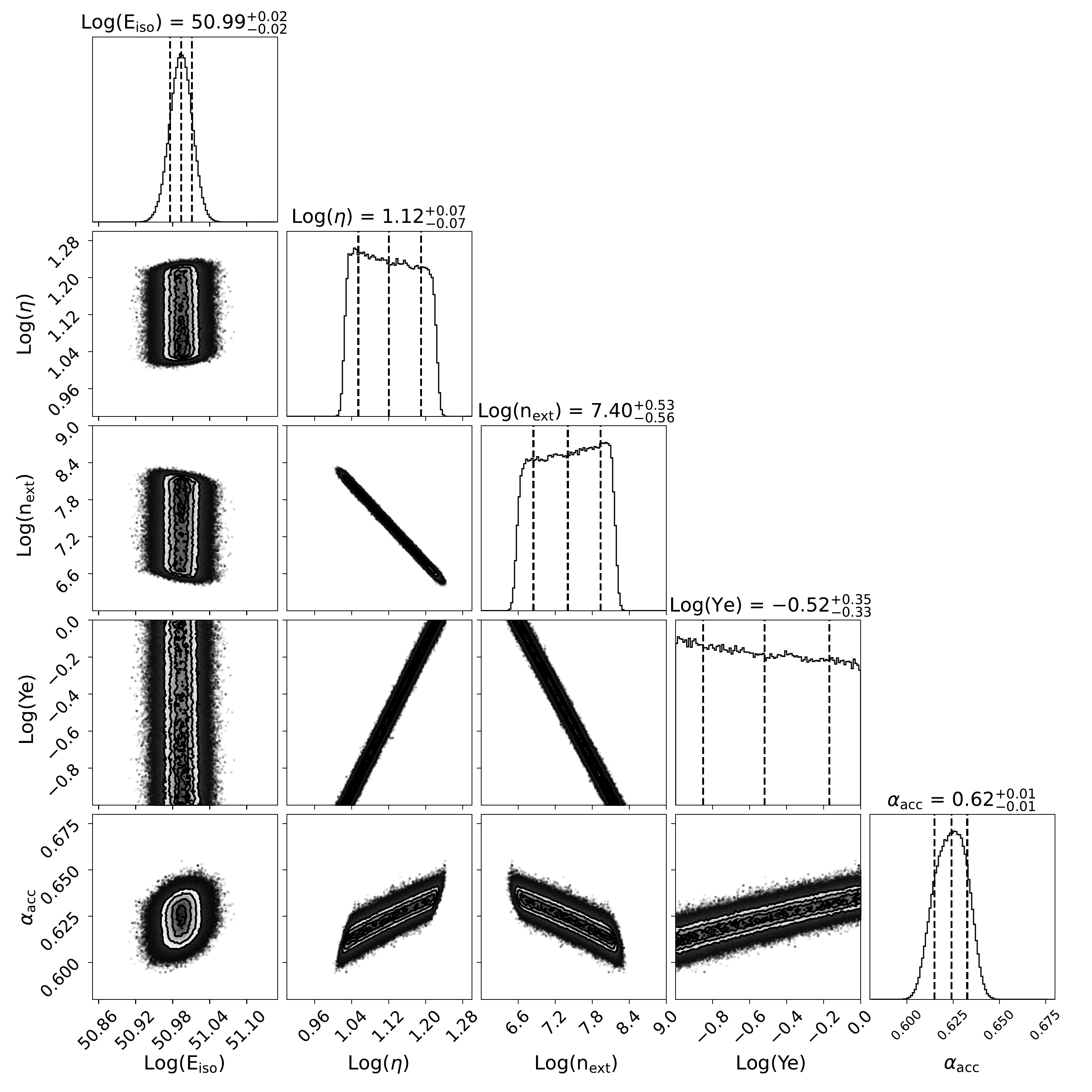}
\caption{Monte Carlo Markov chain parameter estimation for the prompt
light curve of GRB 191019A. All parameters are well constrained, with the exception of 
the electron fraction $Y_e$, which only shows a mild preference for small values.}
\label{fig:mcmc}
\end{figure*}

As described in Section~\ref{sec:methods}, the prompt light curve of GRB~191019A was modeled with five free parameters. We used the Monte Carlo Markov chain implementation emcee \citep{Foreman2013} for minimizing the $\chi^2$.
Flat priors were assumed for the logarithm of the burst isotropic equivalent energy $E_{\rm{iso}}$, the fireball's asymptotic Lorentz factor $\eta$, the external medium density $n_{\rm{ext}}$ and the 
electron fraction $Y_e$. A flat prior in linear space was instead assumed for the acceleration efficiency $\alpha_{\rm{acc}}$. The best fit light curve is shown in Figure~\ref{fig:bestfit} while the corner plot of parameter estimation with posterior uncertainties is shown in Figure~\ref{fig:mcmc}.

The overall fit is acceptable, even if some deviations are noticeable (Figure~\ref{fig:bestfit}). These are due to the fact that we are not attempting to carry out a formal fit to the data. The unknown properties of the GRB engine are too many. Investigating, e.g., whether the engine duration is different from 1.1~s, or whether the number of shells ejected is truly 6 is beyond the scope of this paper. What we focus here is showing that a vanilla short GRB in terms of energy, engine duration and Lorentz factor, can reproduce the overall shape of the burst, and we can set limits on the properties of the external medium. In addition we
note that we have binned the prompt light curve from \cite{Levan2023} and may have smoothed out some short time-scale variability in the process. Some variability on time scales shorter than the overall burst duration is expected in our model, due to the fact that each shell from the engine produces a reverse shock with its individual peak time and duration. As discussed above, a detailed fit with individual shell ejection times, Lorentz factors, thicknesses, and energetics is beyond the scope of this manuscript. In addition, such a fit would not be supported by the low signal to noise ratio of the prompt light curve. We stress, however, that if well-separated peaks were present in the light curve, it would make it difficult to reconcile the data with the model.

As seen in Figure~\ref{fig:mcmc}, a vanilla short GRB is indeed what is preferred by the data, with the exception of the asymptotic Lorentz factor $\eta$ which is found to be small, of order 10. While this is lower than the fiducial on-axis burst expectation ($\eta\sim100$), it is not surprising. The asymptotic Lorentz factor is a function of the viewing angle and the likelihood of seeing a perfectly on-axis burst is vanishing small. Comparing our inferred value with simulation results for the well studied GW~170817 \citep{Lazzati2018,Salafia2020}, we find that a viewing angle of $\sim10^\circ$ is inferred.

Most interesting is the fairly tight constraint on the external density, which is found to be bound by $10^7~{\rm cm}^{-3} \lesssim n_{\rm ext} \lesssim 10^8~{\rm cm}^{-3}$, with a slight preference for the higher values. This is the highest density inferred for any burst to date. A closer inspection of Figure~\ref{fig:mcmc} reveals a tight correlation between external density, Lorentz factor, and electron fraction. Of the three, the electron fraction $Y_e$ is not constrained by the data, which only show a marginal preference for low values over the range allowed by the flat prior. The acceleration efficiency is found to be fairly low, in keeping with numerical predictions of short GRB jets expanding in the dynamical ejecta of the binary \citep{Murguia2014,Nagakura2014,Lazzati2018}.

The spectral information from the model was not used since the spectrum of GRB~191019A is consistent with a single steep power-law, indicating that the peak frequency lies below the instrumental band, or near its edge. We carried out an a-posteriori check and found that our best model predicts a peak frequency of approximately 3 keV, consistent with the observational constraints.

\section{Summary and discussion}
\label{discussion}

We have shown that GRB~191019A can be well modeled as a short GRB of intrinsic duration 1.1~s, isotropic energy $E_{\rm iso}=10^{51}$~erg, and moderate Lorentz factor $\eta=10$,
emerging from a circumburst medium of density $n_{\rm ext}\sim 10^7-10^8$~cm$^{-3}$. This value is higher than even the densest known  molecular clouds, but fully consistent with the outer parts of an AGN disk.

 The host galaxy	of GRB~191019A	has a smooth morphology with a very
compact, almost point-like core. A scaling relation with the galaxy
mass would imply a central SMBH of a few $\times 10^7
M_\odot$. \citet{Levan2023} explored the possibility that the central
light concentration could arise from an AGN.  The observation of a
weak [N$_{\rm II}$] line combined with the absence of hydrogen or oxygen
lines was found to be compatible with an AGN-like set of line ratios,
although this is not conclusive. Late observations with the {\em Swift} X-ray
Telescope set a limit on the X-ray luminosity of $L_X<6\times
10^{42}$~erg~s$^{-1}$, which rules out a high-luminosity AGN, while
still leaving the possibility of a low-luminosity
one\footnote{Measurements of the X-ray luminosity function of AGNs in
the low-redshift Universe \citep{Ueda2003} show that the low
luminosity AGNs ($\sim 10^{41}-10^{42}$~erg~s$^{-1}$) are largely
outnumbering the high luminosity ones.}.
The localization of GRB~191019A	within the innermost 100~pc of its
host galaxy makes the AGN disk interpretation especially tantalizing.
However we note	that, for this scenario	to hold, either	the SMBH mass
has to be somewhat lower than the relatively high value implied	by the
galaxy mass, or	accretion has to be somewhat lower than the close-to-Eddington rate typically assumed, to allow for both
a low-luminosity AGN as	well as	a not too massive disk that would diffuse the GRB radiation (see discussion
in \citealt{Perna2021})\footnote{The latter condition can however be somewhat relaxed considering the
fact that GRB jets can excavate a funnel in the disk and their cocoon emerge
\citep{Tagawa2022}.}. In the following,	to make	our discussion
more quantitative using specific	examples from the literature, we will
consider an AGN	disk around a SMBH of $4\times 10^6 M_\odot$.

In order to relate the inferred circumburst density to that of an AGN disk, we can refer to specific disk models which have been developed in the literature, and in particular the ones of \citet{Sirko2003} and of \citet{Thompson2005}. A central density of $\sim 10^7-10^8$~cm$^{-3}$ is realized between $\sim 5\times 10^5-10^6$ gravitational radii ($R_g$) for the former model, and  $\sim (10^6-3\times 10^6)~R_g$ for the latter. 
For a $4\times 10^6 M_\odot$ SMBH, the range of density inferred for the ambient medium of GRB~101019A would hence correspond to radial distances between a tenth of pc to a pc.

It is interesting to compare these findings with results of numerical simulations of a population of neutron stars (NSs) evolving in the specific environment of an AGN disk. Using the $N$-body  code developed by \citet{Tagawa2020} for compact object evolution in an AGN disk, \citet{Perna2021b} investigated the fate of the NS population. A generic finding of their modeling (i.e.\@ independent of the model parameters) is that NS-NS mergers preferentially occur in two regions within the disk, one at very small radii, and another at large radii. 
For their fiducial model with SMBH mass $4\times 10^6~M_\odot$,
they found these two regions
to be located at about $\sim 10^{-4}-10^{-3}$~pc and 
$\sim 0.1-1$~pc, respectively. The physical reason for this bimodal distribution lies in the fact that, below $\sim 0.1$~pc, migration times are fast enough for the NSs to migrate towards the central disk regions within the AGN lifetime. On the other hand, NSs initially in the outer disk regions, where migration times are very long, will tend to remain in the same sites where they were either born or captured from the nuclear star cluster.
Therefore, theoretical models predict a roughly bimodal NS-NS merger distribution (see Fig.2 in \citealt{Perna2021b}). It is especially intriguing the fact that the density in the outer disk regions that we infer from the best fit modeling to the light curve of GRB~191019A points to one of the two regions (namely the outer one) from which NS-NS mergers are in fact expected with higher probability. In addition, these regions are also those in which the burst is
expected to emerge from the disk without being diffused or significantly absorbed by the disk material (see Figure 5 in \citealt{Perna2021}), consistent with the fact that the burst appearance is intrinsic within our model. While the disk has a significant column density for the considered scenario ($N_H\sim10^{22}$~cm$^{-2}$, assuming a disk thickness $\sim1\%$ of the radius), the burst itself would destroy dust and photoionize the whole disk material, rendering any absorption undetectable \citep{Ray2023}. This is also consistent with GRB~191019A not showing significant absorption in its afterglow \citep{Levan2023}.

Our modeling implies a modest initial Lorentz factor $\eta \sim 10$, suggestive of a viewing angle $\sim 10^\circ$ \citep{Lazzati2018,Salafia2020}. 
The NS-NS binary is likely to have its orbital angular momentum aligned with that of the AGN disk (due to either birth within the disk itself, or orbital alignment after capture from the nuclear star cluster\footnote{The former mechanism is dominant in the outer disk regions due to the long alignment times for capture in those regions, see e.g. \citet{Fabj2020}.}), unless a scattering with a tertiary perturbs its orbit \citep{Samsing2022}. 
The jet of a short GRB is in turn likely pointing in the direction of the orbital angular momentum (e.g. \citealt{Ruiz2016}). Therefore, our inferred viewing angle of $10^\circ$ is likely indicating the inclination of the host disk with respect to the observer.
The angular momentum of the disk is hence misaligned with respect to the angular momentum of the galaxy, which is seen nearly edge on \citep{Levan2023}.
This is not surprising, given that 
observations of various kinds point to a near complete lack of correlation between these two quantities (e.g. \citealt{kinney2000}), and numerical simulations (e.g. \citealt{Hopkins2012}) further support such misalignment.
Our interpretation and modeling of the event GRB~191019A
hence leads to a yet novel way to measure the AGN disk/galaxy offset.

The only part of the data that we do not include in our model is the afterglow. GRB~191019A had an X-ray afterglow and a single detection in the optical \citep{Levan2023}. Given the burst redshift, the afterglow is weaker than average but not particularly unusual. A precise modeling of GRB afterglows in dense media is made difficult by the fact that the self-absorption frequency can exceed the injection frequency, possibly preventing the formation of a power-law distribution \citep{Ghisellini1998}. For these reasons we only notice that a general prediction of a dimmer than average afterglow is consistent with the model characteristics \citep{Lazzati2022} but we do not attempt a formal afterglow model. A future fit of the aftrerglow data when a suitable model is developed may offer further support to our interpretation. We finally notice that   GRB~191019A is not the only burst with long duration detected at low redshift and lacking a SN component. In all other cases (GRB~060505, \citealt{Fynbo2006}; GRB~060614, \citealt{DellaValla2006,Gal-Yam2006,Gehrels2006}; GRB~211211A, \citealt{Mei2022,Rastinejad2022,Troja2022,Yang2022}), however, a clear offset from the host center was detected, making the  case of GRB~191019A unique. In addition the longer of these bursts, GRB~060614 and GRB~211211A, had complex, multi-peaked prompt light curves \citep[e.g.][]{Gompertz23}, differently from GRB~191019A. This underlines that there may be multiple mechanisms by which a short burst engine can produce bursts with prompt gamma-ray emission lasting longer than the canonical 2~s.

\software{Python (https://www.python.org/), emcee \citep{Foreman2013}}

\acknowledgements We would like to acknowledge insightful comments from Daniele Bj{\o}rn Malesani and by the anonymous referee that helped improving the manuscript. DL acknowledges support from NSF grant AST-1907955.
RP acknowledges support by NSF grant AST-2006839.

%

\bibliographystyle{aasjournal}
\bibliography{biblio}

\begin{thebibliography}{}
\expandafter\ifx\csname natexlab\endcsname\relax\def\natexlab#1{#1}\fi
\providecommand{\url}[1]{\href{#1}{#1}}
\providecommand{\dodoi}[1]{doi:~\href{http://doi.org/#1}{\nolinkurl{#1}}}
\providecommand{\doeprint}[1]{\href{http://ascl.net/#1}{\nolinkurl{http://ascl.net/#1}}}
\providecommand{\doarXiv}[1]{\href{https://arxiv.org/abs/#1}{\nolinkurl{https://arxiv.org/abs/#1}}}

\bibitem[{{Abbott} {et~al.}(2017){Abbott}, {Abbott}, {Abbott}, {Acernese},
  {Ackley}, {Adams}, {Adams}, {Addesso}, {Adhikari}, {Adya}, {Affeldt},
  {Afrough}, {Agarwal}, {Agathos}, {Agatsuma}, {Aggarwal}, {Aguiar}, {Aiello},
  {Ain}, {Ajith}, {Allen}, {Allen}, {Allocca}, {Altin}, {Amato}, {Ananyeva},
  {Anderson}, {Anderson}, {Angelova}, {Antier}, {Appert}, {Arai}, {Araya},
  {Areeda}, {Arnaud}, {Arun}, {Ascenzi}, {Ashton}, {Ast}, {Aston}, {Astone},
  {Atallah}, {Aufmuth}, {Aulbert}, {AultONeal}, {Austin}, {Avila-Alvarez},
  {Babak}, {Bacon}, {Bader}, {Bae}, {Baker}, {Baldaccini}, {Ballardin},
  {Ballmer}, {Banagiri}, {Barayoga}, {Barclay}, {Barish}, {Barker}, {Barkett},
  {Barone}, {Barr}, {Barsotti}, {Barsuglia}, {Barta}, {Barthelmy}, {Bartlett},
  {Bartos}, {Bassiri}, {Basti}, {Batch}, {Bawaj}, {Bayley}, {Bazzan},
  {B{\'e}csy}, {Beer}, {Bejger}, {Belahcene}, {Bell}, {Berger}, {Bergmann},
  {Bero}, {Berry}, {Bersanetti}, {Bertolini}, {Betzwieser}, {Bhagwat},
  {Bhandare}, {Bilenko}, {Billingsley}, {Billman}, {Birch}, {Birney},
  {Birnholtz}, {Biscans}, {Biscoveanu}, {Bisht}, {Bitossi}, {Biwer},
  {Bizouard}, {Blackburn}, {Blackman}, {Blair}, {Blair}, {Blair}, {Bloemen},
  {Bock}, {Bode}, {Boer}, {Bogaert}, {Bohe}, {Bondu}, {Bonilla}, {Bonnand},
  {Boom}, {Bork}, {Boschi}, {Bose}, {Bossie}, {Bouffanais}, {Bozzi},
  {Bradaschia}, {Brady}, {Branchesi}, {Brau}, {Briant}, {Brillet}, {Brinkmann},
  {Brisson}, {Brockill}, {Broida}, {Brooks}, {Brown}, {Brown}, {Brunett},
  {Buchanan}, {Buikema}, {Bulik}, {Bulten}, {Buonanno}, {Buskulic}, {Buy},
  {Byer}, {Cabero}, {Cadonati}, {Cagnoli}, {Cahillane}, {Calder{\'o}n
  Bustillo}, {Callister}, {Calloni}, {Camp}, {Canepa}, {Canizares}, {Cannon},
  {Cao}, {Cao}, {Capano}, {Capocasa}, {Carbognani}, {Caride}, {Carney},
  {Casanueva Diaz}, {Casentini}, {Caudill}, {Cavagli{\`a}}, {Cavalier},
  {Cavalieri}, {Cella}, {Cepeda}, {Cerd{\'a}-Dur{\'a}n}, {Cerretani},
  {Cesarini}, {Chamberlin}, {Chan}, {Chao}, {Charlton}, {Chase},
  {Chassande-Mottin}, {Chatterjee}, {Chatziioannou}, {Cheeseboro}, {Chen},
  {Chen}, {Chen}, {Cheng}, {Chia}, {Chincarini}, {Chiummo}, {Chmiel}, {Cho},
  {Cho}, {Chow}, {Christensen}, {Chu}, {Chua}, {Chua}, {Chung}, {Chung},
  {Ciani}, {Ciolfi}, {Cirelli}, {Cirone}, {Clara}, {Clark}, {Clearwater},
  {Cleva}, {Cocchieri}, {Coccia}, {Cohadon}, {Cohen}, {Colla}, {Collette},
  {Cominsky}, {Constancio}, {Conti}, {Cooper}, {Corban}, {Corbitt},
  {Cordero-Carri{\'o}n}, {Corley}, {Cornish}, {Corsi}, {Cortese}, {Costa},
  {Coughlin}, {Coughlin}, {Coulon}, {Countryman}, {Couvares}, {Covas}, {Cowan},
  {Coward}, {Cowart}, {Coyne}, {Coyne}, {Creighton}, {Creighton}, {Cripe},
  {Crowder}, {Cullen}, {Cumming}, {Cunningham}, {Cuoco}, {Dal Canton},
  {D{\'a}lya}, {Danilishin}, {D'Antonio}, {Danzmann}, {Dasgupta}, {Da Silva
  Costa}, {Dattilo}, {Dave}, {Davier}, {Davis}, {Daw}, {Day}, {De}, {DeBra},
  {Degallaix}, {De Laurentis}, {Del{\'e}glise}, {Del Pozzo}, {Demos}, {Denker},
  {Dent}, {De Pietri}, {Dergachev}, {De Rosa}, {DeRosa}, {De Rossi}, {DeSalvo},
  {de Varona}, {Devenson}, {Dhurandhar}, {D{\'\i}az}, {Di Fiore}, {Di
  Giovanni}, {Di Girolamo}, {Di Lieto}, {Di Pace}, {Di Palma}, {Di Renzo},
  {Doctor}, {Dolique}, {Donovan}, {Dooley}, {Doravari}, {Dorrington},
  {Douglas}, {Dovale {\'A}lvarez}, {Downes}, {Drago}, {Dreissigacker},
  {Driggers}, {Du}, {Ducrot}, {Dupej}, {Dwyer}, {Edo}, {Edwards}, {Effler},
  {Ehrens}, {Eichholz}, {Eikenberry}, {Eisenstein}, {Essick}, {Estevez},
  {Etienne}, {Etzel}, {Evans}, {Evans}, {Factourovich}, {Fafone}, {Fair},
  {Fairhurst}, {Fan}, {Farinon}, {Farr}, {Farr}, {Fauchon-Jones}, {Favata},
  {Fays}, {Fee}, {Fehrmann}, {Feicht}, {Fejer}, {Fernandez-Galiana},
  {Ferrante}, {Ferreira}, {Ferrini}, {Fidecaro}, {Finstad}, {Fiori},
  {Fiorucci}, {Fishbach}, {Fisher}, {Fitz-Axen}, {Flaminio}, {Fletcher},
  {Fong}, {Font}, {Forsyth}, {Forsyth}, {Fournier}, {Frasca}, {Frasconi},
  {Frei}, {Freise}, {Frey}, {Frey}, {Fries}, {Fritschel}, {Frolov}, {Fulda},
  {Fyffe}, {Gabbard}, {Gadre}, {Gaebel}, {Gair}, {Gammaitoni}, {Ganija},
  {Gaonkar}, {Garcia-Quiros}, {Garufi}, {Gateley}, {Gaudio}, {Gaur},
  {Gayathri}, {Gehrels}, {Gemme}, {Genin}, {Gennai}, {George}, {George},
  {Gergely}, {Germain}, {Ghonge}, {Ghosh}, {Ghosh}, {Ghosh}, {Giaime},
  {Giardina}, {Giazotto}, {Gill}, {Glover}, {Goetz}, {Goetz}, {Gomes},
  {Goncharov}, {Gonz{\'a}lez}, {Gonzalez Castro}, {Gopakumar}, {Gorodetsky},
  {Gossan}, {Gosselin}, {Gouaty}, {Grado}, {Graef}, {Granata}, {Grant}, {Gras},
  {Gray}, {Greco}, {Green}, {Gretarsson}, {Griswold}, {Groot}, {Grote},
  {Grunewald}, {Gruning}, {Guidi}, {Guo}, {Gupta}, {Gupta}, {Gushwa},
  {Gustafson}, {Gustafson}, {Halim}, {Hall}, {Hall}, {Hamilton}, {Hammond},
  {Haney}, {Hanke}, {Hanks}, {Hanna}, {Hannam}, {Hannuksela}, {Hanson},
  {Hardwick}, {Harms}, {Harry}, {Harry}, {Hart}, {Haster}, {Haughian}, {Healy},
  {Heidmann}, {Heintze}, {Heitmann}, {Hello}, {Hemming}, {Hendry}, {Heng},
  {Hennig}, {Heptonstall}, {Heurs}, {Hild}, {Hinderer}, {Hoak}, {Hofman},
  {Holt}, {Holz}, {Hopkins}, {Horst}, {Hough}, {Houston}, {Howell}, {Hreibi},
  {Hu}, {Huerta}, {Huet}, {Hughey}, {Husa}, {Huttner}, {Huynh-Dinh}, {Indik},
  {Inta}, {Intini}, {Isa}, {Isac}, {Isi}, {Iyer}, {Izumi}, {Jacqmin}, {Jani},
  {Jaranowski}, {Jawahar}, {Jim{\'e}nez-Forteza}, {Johnson}, {Jones}, {Jones},
  {Jonker}, {Ju}, {Junker}, {Kalaghatgi}, {Kalogera}, {Kamai}, {Kandhasamy},
  {Kang}, {Kanner}, {Kapadia}, {Karki}, {Karvinen}, {Kasprzack}, {Katolik},
  {Katsavounidis}, {Katzman}, {Kaufer}, {Kawabe}, {K{\'e}f{\'e}lian}, {Keitel},
  {Kemball}, {Kennedy}, {Kent}, {Key}, {Khalili}, {Khan}, {Khan}, {Khan},
  {Khazanov}, {Kijbunchoo}, {Kim}, {Kim}, {Kim}, {Kim}, {Kim}, {Kim},
  {Kimbrell}, {King}, {King}, {Kinley-Hanlon}, {Kirchhoff}, {Kissel},
  {Kleybolte}, {Klimenko}, {Knowles}, {Koch}, {Koehlenbeck}, {Koley},
  {Kondrashov}, {Kontos}, {Korobko}, {Korth}, {Kowalska}, {Kozak},
  {Kr{\"a}mer}, {Kringel}, {Krishnan}, {Kr{\'o}lak}, {Kuehn}, {Kumar}, {Kumar},
  {Kumar}, {Kuo}, {Kutynia}, {Kwang}, {Lackey}, {Lai}, {Landry}, {Lang},
  {Lange}, {Lantz}, {Lanza}, {Larson}, {Lartaux-Vollard}, {Lasky}, {Laxen},
  {Lazzarini}, {Lazzaro}, {Leaci}, {Leavey}, {Lee}, {Lee}, {Lee}, {Lee}, {Lee},
  {Lehmann}, {Lenon}, {Leonardi}, {Leroy}, {Letendre}, {Levin}, {Li}, {Linker},
  {Littenberg}, {Liu}, {Lo}, {Lockerbie}, {London}, {Lord}, {Lorenzini},
  {Loriette}, {Lormand}, {Losurdo}, {Lough}, {Lousto}, {Lovelace}, {L{\"u}ck},
  {Lumaca}, {Lundgren}, {Lynch}, {Ma}, {Macas}, {Macfoy}, {Machenschalk},
  {MacInnis}, {Macleod}, {Maga{\~n}a Hernandez}, {Maga{\~n}a-Sandoval},
  {Maga{\~n}a Zertuche}, {Magee}, {Majorana}, {Maksimovic}, {Man}, {Mandic},
  {Mangano}, {Mansell}, {Manske}, {Mantovani}, {Marchesoni}, {Marion},
  {M{\'a}rka}, {M{\'a}rka}, {Markakis}, {Markosyan}, {Markowitz}, {Maros},
  {Marquina}, {Marsh}, {Martelli}, {Martellini}, {Martin}, {Martin},
  {Martynov}, {Mason}, {Massera}, {Masserot}, {Massinger}, {Masso-Reid},
  {Mastrogiovanni}, {Matas}, {Matichard}, {Matone}, {Mavalvala}, {Mazumder},
  {McCarthy}, {McClelland}, {McCormick}, {McCuller}, {McGuire}, {McIntyre},
  {McIver}, {McManus}, {McNeill}, {McRae}, {McWilliams}, {Meacher}, {Meadors},
  {Mehmet}, {Meidam}, {Mejuto-Villa}, {Melatos}, {Mendell}, {Mercer}, {Merilh},
  {Merzougui}, {Meshkov}, {Messenger}, {Messick}, {Metzdorff}, {Meyers},
  {Miao}, {Michel}, {Middleton}, {Mikhailov}, {Milano}, {Miller}, {Miller},
  {Miller}, {Millhouse}, {Milovich-Goff}, {Minazzoli}, {Minenkov}, {Ming},
  {Mishra}, {Mitra}, {Mitrofanov}, {Mitselmakher}, {Mittleman}, {Moffa},
  {Moggi}, {Mogushi}, {Mohan}, {Mohapatra}, {Montani}, {Moore}, {Moraru},
  {Moreno}, {Morriss}, {Mours}, {Mow-Lowry}, {Mueller}, {Muir}, {Mukherjee},
  {Mukherjee}, {Mukherjee}, {Mukund}, {Mullavey}, {Munch}, {Mu{\~n}iz},
  {Muratore}, {Murray}, {Napier}, {Nardecchia}, {Naticchioni}, {Nayak},
  {Neilson}, {Nelemans}, {Nelson}, {Nery}, {Neunzert}, {Nevin}, {Newport},
  {Newton}, {Ng}, {Nguyen}, {Nguyen}, {Nichols}, {Nielsen}, {Nissanke}, {Nitz},
  {Noack}, {Nocera}, {Nolting}, {North}, {Nuttall}, {Oberling}, {O'Dea},
  {Ogin}, {Oh}, {Oh}, {Ohme}, {Okada}, {Oliver}, {Oppermann}, {Oram},
  {O'Reilly}, {Ormiston}, {Ortega}, {O'Shaughnessy}, {Ossokine}, {Ottaway},
  {Overmier}, {Owen}, {Pace}, {Page}, {Page}, {Pai}, {Pai}, {Palamos},
  {Palashov}, {Palomba}, {Pal-Singh}, {Pan}, {Pan}, {Pang}, {Pang}, {Pankow},
  {Pannarale}, {Pant}, {Paoletti}, {Paoli}, {Papa}, {Parida}, {Parker},
  {Pascucci}, {Pasqualetti}, {Passaquieti}, {Passuello}, {Patil}, {Patricelli},
  {Pearlstone}, {Pedraza}, {Pedurand}, {Pekowsky}, {Pele}, {Penn}, {Perez},
  {Perreca}, {Perri}, {Pfeiffer}, {Phelps}, {Piccinni}, {Pichot},
  {Piergiovanni}, {Pierro}, {Pillant}, {Pinard}, {Pinto}, {Pirello}, {Pitkin},
  {Poe}, {Poggiani}, {Popolizio}, {Porter}, {Post}, {Powell}, {Prasad},
  {Pratt}, {Pratten}, {Predoi}, {Prestegard}, {Price}, {Prijatelj}, {Principe},
  {Privitera}, {Prodi}, {Prokhorov}, {Puncken}, {Punturo}, {Puppo},
  {P{\"u}rrer}, {Qi}, {Quetschke}, {Quintero}, {Quitzow-James}, {Raab},
  {Rabeling}, {Radkins}, {Raffai}, {Raja}, {Rajan}, {Rajbhandari}, {Rakhmanov},
  {Ramirez}, {Ramos-Buades}, {Rapagnani}, {Raymond}, {Razzano}, {Read},
  {Regimbau}, {Rei}, {Reid}, {Reitze}, {Ren}, {Reyes}, {Ricci}, {Ricker},
  {Rieger}, {Riles}, {Rizzo}, {Robertson}, {Robie}, {Robinet}, {Rocchi},
  {Rolland}, {Rollins}, {Roma}, {Romano}, {Romel}, {Romie}, {Rosi{\'n}ska},
  {Ross}, {Rowan}, {R{\"u}diger}, {Ruggi}, {Rutins}, {Ryan}, {Sachdev},
  {Sadecki}, {Sadeghian}, {Sakellariadou}, {Salconi}, {Saleem}, {Salemi},
  {Samajdar}, {Sammut}, {Sampson}, {Sanchez}, {Sanchez}, {Sanchis-Gual},
  {Sandberg}, {Sanders}, {Sassolas}, {Sathyaprakash}, {Saulson}, {Sauter},
  {Savage}, {Sawadsky}, {Schale}, {Scheel}, {Scheuer}, {Schmidt}, {Schmidt},
  {Schnabel}, {Schofield}, {Sch{\"o}nbeck}, {Schreiber}, {Schuette}, {Schulte},
  {Schutz}, {Schwalbe}, {Scott}, {Scott}, {Seidel}, {Sellers}, {Sengupta},
  {Sentenac}, {Sequino}, {Sergeev}, {Shaddock}, {Shaffer}, {Shah}, {Shahriar},
  {Shaner}, {Shao}, {Shapiro}, {Shawhan}, {Sheperd}, {Shoemaker}, {Shoemaker},
  {Siellez}, {Siemens}, {Sieniawska}, {Sigg}, {Silva}, {Singer}, {Singh},
  {Singhal}, {Sintes}, {Slagmolen}, {Smith}, {Smith}, {Smith}, {Somala}, {Son},
  {Sonnenberg}, {Sorazu}, {Sorrentino}, {Souradeep}, {Spencer}, {Srivastava},
  {Staats}, {Staley}, {Steinke}, {Steinlechner}, {Steinlechner}, {Steinmeyer},
  {Stevenson}, {Stone}, {Stops}, {Strain}, {Stratta}, {Strigin}, {Strunk},
  {Sturani}, {Stuver}, {Summerscales}, {Sun}, {Sunil}, {Suresh}, {Sutton},
  {Swinkels}, {Szczepa{\'n}czyk}, {Tacca}, {Tait}, {Talbot}, {Talukder},
  {Tanner}, {T{\'a}pai}, {Taracchini}, {Tasson}, {Taylor}, {Taylor}, {Tewari},
  {Theeg}, {Thies}, {Thomas}, {Thomas}, {Thomas}, {Thorne}, {Thorne}, {Thrane},
  {Tiwari}, {Tiwari}, {Tokmakov}, {Toland}, {Tonelli}, {Tornasi},
  {Torres-Forn{\'e}}, {Torrie}, {T{\"o}yr{\"a}}, {Travasso}, {Traylor},
  {Trinastic}, {Tringali}, {Trozzo}, {Tsang}, {Tse}, {Tso}, {Tsukada}, {Tsuna},
  {Tuyenbayev}, {Ueno}, {Ugolini}, {Unnikrishnan}, {Urban}, {Usman},
  {Vahlbruch}, {Vajente}, {Valdes}, {van Bakel}, {van Beuzekom}, {van den
  Brand}, {Van Den Broeck}, {Vander-Hyde}, {van der Schaaf}, {van Heijningen},
  {van Veggel}, {Vardaro}, {Varma}, {Vass}, {Vas{\'u}th}, {Vecchio},
  {Vedovato}, {Veitch}, {Veitch}, {Venkateswara}, {Venugopalan}, {Verkindt},
  {Vetrano}, {Vicer{\'e}}, {Viets}, {Vinciguerra}, {Vine}, {Vinet}, {Vitale},
  {Vo}, {Vocca}, {Vorvick}, {Vyatchanin}, {Wade}, {Wade}, {Wade}, {Walet},
  {Walker}, {Wallace}, {Walsh}, {Wang}, {Wang}, {Wang}, {Wang}, {Wang}, {Ward},
  {Warner}, {Was}, {Watchi}, {Weaver}, {Wei}, {Weinert}, {Weinstein}, {Weiss},
  {Wen}, {Wessel}, {Wessels}, {Westerweck}, {Westphal}, {Wette}, {Whelan},
  {Whitcomb}, {Whiting}, {Whittle}, {Wilken}, {Williams}, {Williams},
  {Williamson}, {Willis}, {Willke}, {Wimmer}, {Winkler}, {Wipf}, {Wittel},
  {Woan}, {Woehler}, {Wofford}, {Wong}, {Worden}, {Wright}, {Wu}, {Wysocki},
  {Xiao}, {Yamamoto}, {Yancey}, {Yang}, {Yap}, {Yazback}, {Yu}, {Yu}, {Yvert},
  {Zadro{\.z}ny}, {Zanolin}, {Zelenova}, {Zendri}, {Zevin}, {Zhang}, {Zhang},
  {Zhang}, {Zhang}, {Zhao}, {Zhou}, {Zhou}, {Zhu}, {Zhu}, {Zimmerman},
  {Zucker}, {Zweizig}, {LIGO Scientific Collaboration}, {Virgo Collaboration},
  {Wilson-Hodge}, {Bissaldi}, {Blackburn}, {Briggs}, {Burns}, {Cleveland},
  {Connaughton}, {Gibby}, {Giles}, {Goldstein}, {Hamburg}, {Jenke}, {Hui},
  {Kippen}, {Kocevski}, {McBreen}, {Meegan}, {Paciesas}, {Poolakkil}, {Preece},
  {Racusin}, {Roberts}, {Stanbro}, {Veres}, {von Kienlin}, {GBM}, {Savchenko},
  {Ferrigno}, {Kuulkers}, {Bazzano}, {Bozzo}, {Brandt}, {Chenevez},
  {Courvoisier}, {Diehl}, {Domingo}, {Hanlon}, {Jourdain}, {Laurent}, {Lebrun},
  {Lutovinov}, {Martin-Carrillo}, {Mereghetti}, {Natalucci}, {Rodi}, {Roques},
  {Sunyaev}, {Ubertini}, {INTEGRAL}, {Aartsen}, {Ackermann}, {Adams},
  {Aguilar}, {Ahlers}, {Ahrens}, {Samarai}, {Altmann}, {Andeen}, {Anderson},
  {Ansseau}, {Anton}, {Arg{\"u}elles}, {Auffenberg}, {Axani}, {Bagherpour},
  {Bai}, {Barron}, {Barwick}, {Baum}, {Bay}, {Beatty}, {Becker Tjus},
  {Bernardini}, {Besson}, {Binder}, {Bindig}, {Blaufuss}, {Blot}, {Bohm},
  {B{\"o}rner}, {Bos}, {Bose}, {B{\"o}ser}, {Botner}, {Bourbeau}, {Bourbeau},
  {Bradascio}, {Braun}, {Brayeur}, {Brenzke}, {Bretz}, {Bron},
  {Brostean-Kaiser}, {Burgman}, {Carver}, {Casey}, {Casier}, {Cheung},
  {Chirkin}, {Christov}, {Clark}, {Classen}, {Coenders}, {Collin}, {Conrad},
  {Cowen}, {Cross}, {Day}, {de Andr{\'e}}, {De Clercq}, {DeLaunay},
  {Dembinski}, {De Ridder}, {Desiati}, {de Vries}, {de Wasseige}, {de With},
  {DeYoung}, {D{\'\i}az-V{\'e}lez}, {di Lorenzo}, {Dujmovic}, {Dumm},
  {Dunkman}, {Dvorak}, {Eberhardt}, {Ehrhardt}, {Eichmann}, {Eller}, {Evenson},
  {Fahey}, {Fazely}, {Felde}, {Filimonov}, {Finley}, {Flis}, {Franckowiak},
  {Friedman}, {Fuchs}, {Gaisser}, {Gallagher}, {Gerhardt}, {Ghorbani}, {Giang},
  {Glauch}, {Gl{\"u}senkamp}, {Goldschmidt}, {Gonzalez}, {Grant}, {Griffith},
  {Haack}, {Hallgren}, {Halzen}, {Hanson}, {Hebecker}, {Heereman}, {Helbing},
  {Hellauer}, {Hickford}, {Hignight}, {Hill}, {Hoffman}, {Hoffmann},
  {Hokanson-Fasig}, {Hoshina}, {Huang}, {Huber}, {Hultqvist}, {H{\"u}nnefeld},
  {In}, {Ishihara}, {Jacobi}, {Japaridze}, {Jeong}, {Jero}, {Jones},
  {Kalaczynski}, {Kang}, {Kappes}, {Karg}, {Karle}, {Kauer}, {Keivani},
  {Kelley}, {Kheirandish}, {Kim}, {Kim}, {Kintscher}, {Kiryluk}, {Kittler},
  {Klein}, {Kohnen}, {Koirala}, {Kolanoski}, {K{\"o}pke}, {Kopper}, {Kopper},
  {Koschinsky}, {Koskinen}, {Kowalski}, {Krings}, {Kroll}, {Kr{\"u}ckl},
  {Kunnen}, {Kunwar}, {Kurahashi}, {Kuwabara}, {Kyriacou}, {Labare},
  {Lanfranchi}, {Larson}, {Lauber}, {Lesiak-Bzdak}, {Leuermann}, {Liu}, {Lu},
  {L{\"u}nemann}, {Luszczak}, {Madsen}, {Maggi}, {Mahn}, {Mancina}, {Maruyama},
  {Mase}, {Maunu}, {McNally}, {Meagher}, {Medici}, {Meier}, {Menne}, {Merino},
  {Meures}, {Miarecki}, {Micallef}, {Moment{\'e}}, {Montaruli}, {Moore},
  {Moulai}, {Nahnhauer}, {Nakarmi}, {Naumann}, {Neer}, {Niederhausen},
  {Nowicki}, {Nygren}, {Obertacke Pollmann}, {Olivas}, {O'Murchadha},
  {Palczewski}, {Pandya}, {Pankova}, {Peiffer}, {Pepper}, {P{\'e}rez de los
  Heros}, {Pieloth}, {Pinat}, {Price}, {Przybylski}, {Raab}, {R{\"a}del},
  {Rameez}, {Rawlins}, {Rea}, {Reimann}, {Relethford}, {Relich}, {Resconi},
  {Rhode}, {Richman}, {Robertson}, {Rongen}, {Rott}, {Ruhe}, {Ryckbosch},
  {Rysewyk}, {S{\"a}lzer}, {Sanchez Herrera}, {Sandrock}, {Sandroos},
  {Santander}, {Sarkar}, {Sarkar}, {Satalecka}, {Schlunder}, {Schmidt},
  {Schneider}, {Schoenen}, {Sch{\"o}neberg}, {Schumacher}, {Seckel},
  {Seunarine}, {Soedingrekso}, {Soldin}, {Song}, {Spiczak}, {Spiering},
  {Stachurska}, {Stamatikos}, {Stanev}, {Stasik}, {Stettner}, {Steuer},
  {Stezelberger}, {Stokstad}, {St{\"o}ssl}, {Strotjohann}, {Stuttard},
  {Sullivan}, {Sutherland}, {Taboada}, {Tatar}, {Tenholt}, {Ter-Antonyan},
  {Terliuk}, {Te{\v{s}}i{\'c}}, {Tilav}, {Toale}, {Tobin}, {Toscano}, {Tosi},
  {Tselengidou}, {Tung}, {Turcati}, {Turley}, {Ty}, {Unger}, {Usner},
  {Vandenbroucke}, {Van Driessche}, {van Eijndhoven}, {Vanheule}, {van Santen},
  {Vehring}, {Vogel}, {Vraeghe}, {Walck}, {Wallace}, {Wallraff}, {Wandler},
  {Wandkowsky}, {Waza}, {Weaver}, {Weiss}, {Wendt}, {Werthebach}, {Whelan},
  {Wiebe}, {Wiebusch}, {Wille}, {Williams}, {Wills}, {Wolf}, {Wood}, {Woolsey},
  {Woschnagg}, {Xu}, {Xu}, {Xu}, {Yanez}, {Yodh}, {Yoshida}, {Yuan}, {Zoll},
  {IceCube Collaboration}, {Balasubramanian}, {Mate}, {Bhalerao},
  {Bhattacharya}, {Vibhute}, {Dewangan}, {Rao}, {Vadawale}, {AstroSat Cadmium
  Zinc Telluride Imager Team}, {Svinkin}, {Hurley}, {Aptekar}, {Frederiks},
  {Golenetskii}, {Kozlova}, {Lysenko}, {Oleynik}, {Tsvetkova}, {Ulanov},
  {Cline}, {IPN Collaboration}, {Li}, {Xiong}, {Zhang}, {Lu}, {Song}, {Cao},
  {Chang}, {Chen}, {Chen}, {Chen}, {Chen}, {Chen}, {Chen}, {Cui}, {Cui},
  {Deng}, {Dong}, {Du}, {Fu}, {Gao}, {Gao}, {Gao}, {Ge}, {Gu}, {Guan}, {Guo},
  {Han}, {Hu}, {Huang}, {Huo}, {Jia}, {Jiang}, {Jiang}, {Jin}, {Jin}, {Li},
  {Li}, {Li}, {Li}, {Li}, {Li}, {Li}, {Li}, {Li}, {Li}, {Li}, {Liang}, {Liao},
  {Liu}, {Liu}, {Liu}, {Liu}, {Liu}, {Liu}, {Liu}, {Lu}, {Lu}, {Luo}, {Ma},
  {Meng}, {Nang}, {Nie}, {Ou}, {Qu}, {Sai}, {Sun}, {Tan}, {Tao}, {Tao}, {Tuo},
  {Wang}, {Wang}, {Wang}, {Wang}, {Wang}, {Wen}, {Wu}, {Wu}, {Xiao}, {Xu},
  {Xu}, {Yan}, {Yang}, {Yang}, {Yang}, {Zhang}, {Zhang}, {Zhang}, {Zhang},
  {Zhang}, {Zhang}, {Zhang}, {Zhang}, {Zhang}, {Zhang}, {Zhang}, {Zhang},
  {Zhang}, {Zhang}, {Zhang}, {Zhang}, {Zhang}, {Zhang}, {Zhao}, {Zhao}, {Zhao},
  {Zheng}, {Zhu}, {Zhu}, {Zou}, {Insight-HXMT Collaboration}, {Albert},
  {Andr{\'e}}, {Anghinolfi}, {Ardid}, {Aubert}, {Aublin}, {Avgitas}, {Baret},
  {Barrios-Mart{\'\i}}, {Basa}, {Belhorma}, {Bertin}, {Biagi}, {Bormuth},
  {Bourret}, {Bouwhuis}, {Br{\^a}nza{\c{s}}}, {Bruijn}, {Brunner}, {Busto},
  {Capone}, {Caramete}, {Carr}, {Celli}, {Cherkaoui El Moursli}, {Chiarusi},
  {Circella}, {Coelho}, {Coleiro}, {Coniglione}, {Costantini}, {Coyle},
  {Creusot}, {D{\'\i}az}, {Deschamps}, {De Bonis}, {Distefano}, {Di Palma},
  {Domi}, {Donzaud}, {Dornic}, {Drouhin}, {Eberl}, {El Bojaddaini}, {El
  Khayati}, {Els{\"a}sser}, {Enzenh{\"o}fer}, {Ettahiri}, {Fassi}, {Felis},
  {Fusco}, {Gay}, {Giordano}, {Glotin}, {Gr{\'e}goire}, {Ruiz}, {Graf},
  {Hallmann}, {van Haren}, {Heijboer}, {Hello}, {Hern{\'a}ndez-Rey},
  {H{\"o}ssl}, {Hofest{\"a}dt}, {Hugon}, {Illuminati}, {James}, {de Jong},
  {Jongen}, {Kadler}, {Kalekin}, {Katz}, {Kiessling}, {Kouchner}, {Kreter},
  {Kreykenbohm}, {Kulikovskiy}, {Lachaud}, {Lahmann}, {Lef{\`e}vre}, {Leonora},
  {Lotze}, {Loucatos}, {Marcelin}, {Margiotta}, {Marinelli},
  {Mart{\'\i}nez-Mora}, {Mele}, {Melis}, {Michael}, {Migliozzi}, {Moussa},
  {Navas}, {Nezri}, {Organokov}, {P{\u{a}}v{\u{a}}la{\c{s}}}, {Pellegrino},
  {Perrina}, {Piattelli}, {Popa}, {Pradier}, {Quinn}, {Racca}, {Riccobene},
  {S{\'a}nchez-Losa}, {Salda{\~n}a}, {Salvadori}, {Samtleben}, {Sanguineti},
  {Sapienza}, {Sieger}, {Spurio}, {Stolarczyk}, {Taiuti}, {Tayalati},
  {Trovato}, {Turpin}, {T{\"o}nnis}, {Vallage}, {Van Elewyck}, {Versari},
  {Vivolo}, {Vizzoca}, {Wilms}, {Zornoza}, {Z{\'u}{\~n}iga}, {ANTARES
  Collaboration}, {Beardmore}, {Breeveld}, {Burrows}, {Cenko}, {Cusumano},
  {D'A{\`\i}}, {de Pasquale}, {Emery}, {Evans}, {Giommi}, {Gronwall}, {Kennea},
  {Krimm}, {Kuin}, {Lien}, {Marshall}, {Melandri}, {Nousek}, {Oates},
  {Osborne}, {Pagani}, {Page}, {Palmer}, {Perri}, {Siegel}, {Sbarufatti},
  {Tagliaferri}, {Tohuvavohu}, {Swift Collaboration}, {Tavani}, {Verrecchia},
  {Bulgarelli}, {Evangelista}, {Pacciani}, {Feroci}, {Pittori}, {Giuliani},
  {Del Monte}, {Donnarumma}, {Argan}, {Trois}, {Ursi}, {Cardillo}, {Piano},
  {Longo}, {Lucarelli}, {Munar-Adrover}, {Fuschino}, {Labanti}, {Marisaldi},
  {Minervini}, {Fioretti}, {Parmiggiani}, {Gianotti}, {Trifoglio}, {Di Persio},
  {Antonelli}, {Barbiellini}, {Caraveo}, {Cattaneo}, {Costa}, {Colafrancesco},
  {D'Amico}, {Ferrari}, {Morselli}, {Paoletti}, {Picozza}, {Pilia}, {Rappoldi},
  {Soffitta}, {Vercellone}, {AGILE Team}, {Foley}, {Coulter}, {Kilpatrick},
  {Drout}, {Piro}, {Shappee}, {Siebert}, {Simon}, {Ulloa}, {Kasen}, {Madore},
  {Murguia-Berthier}, {Pan}, {Prochaska}, {Ramirez-Ruiz}, {Rest},
  {Rojas-Bravo}, {1M2H Team}, {Berger}, {Soares-Santos}, {Annis}, {Alexander},
  {Allam}, {Balbinot}, {Blanchard}, {Brout}, {Butler}, {Chornock}, {Cook},
  {Cowperthwaite}, {Diehl}, {Drlica-Wagner}, {Drout}, {Durret}, {Eftekhari},
  {Finley}, {Fong}, {Frieman}, {Fryer}, {Garc{\'\i}a-Bellido}, {Gruendl},
  {Hartley}, {Herner}, {Kessler}, {Lin}, {Lopes}, {Louren{\c{c}}o}, {Margutti},
  {Marshall}, {Matheson}, {Medina}, {Metzger}, {Mu{\~n}oz}, {Muir}, {Nicholl},
  {Nugent}, {Palmese}, {Paz-Chinch{\'o}n}, {Quataert}, {Sako}, {Sauseda},
  {Schlegel}, {Scolnic}, {Secco}, {Smith}, {Sobreira}, {Villar}, {Vivas},
  {Wester}, {Williams}, {Yanny}, {Zenteno}, {Zhang}, {Abbott}, {Banerji},
  {Bechtol}, {Benoit-L{\'e}vy}, {Bertin}, {Brooks}, {Buckley-Geer}, {Burke},
  {Capozzi}, {Carnero Rosell}, {Carrasco Kind}, {Castander}, {Crocce}, {Cunha},
  {D'Andrea}, {da Costa}, {Davis}, {DePoy}, {Desai}, {Dietrich}, {Eifler},
  {Fernandez}, {Flaugher}, {Fosalba}, {Gaztanaga}, {Gerdes}, {Giannantonio},
  {Goldstein}, {Gruen}, {Gschwend}, {Gutierrez}, {Honscheid}, {James},
  {Jeltema}, {Johnson}, {Johnson}, {Kent}, {Krause}, {Kron}, {Kuehn}, {Lahav},
  {Lima}, {Maia}, {March}, {Martini}, {McMahon}, {Menanteau}, {Miller},
  {Miquel}, {Mohr}, {Nichol}, {Ogando}, {Plazas}, {Romer}, {Roodman}, {Rykoff},
  {Sanchez}, {Scarpine}, {Schindler}, {Schubnell}, {Sevilla-Noarbe}, {Sheldon},
  {Smith}, {Smith}, {Stebbins}, {Suchyta}, {Swanson}, {Tarle}, {Thomas},
  {Troxel}, {Tucker}, {Vikram}, {Walker}, {Wechsler}, {Weller}, {Carlin},
  {Gill}, {Li}, {Marriner}, {Neilsen}, {Dark Energy Camera GW-EM
  Collaboration}, {DES Collaboration}, {Haislip}, {Kouprianov}, {Reichart},
  {Sand}, {Tartaglia}, {Valenti}, {Yang}, {DLT40 Collaboration}, {Benetti},
  {Brocato}, {Campana}, {Cappellaro}, {Covino}, {D'Avanzo}, {D'Elia}, {Getman},
  {Ghirlanda}, {Ghisellini}, {Limatola}, {Nicastro}, {Palazzi}, {Pian},
  {Piranomonte}, {Possenti}, {Rossi}, {Salafia}, {Tomasella}, {Amati},
  {Antonelli}, {Bernardini}, {Bufano}, {Capaccioli}, {Casella}, {Dadina}, {De
  Cesare}, {Di Paola}, {Giuffrida}, {Giunta}, {Israel}, {Lisi}, {Maiorano},
  {Mapelli}, {Masetti}, {Pescalli}, {Pulone}, {Salvaterra}, {Schipani},
  {Spera}, {Stamerra}, {Stella}, {Testa}, {Turatto}, {Vergani}, {Aresu},
  {Bachetti}, {Buffa}, {Burgay}, {Buttu}, {Caria}, {Carretti}, {Casasola},
  {Castangia}, {Carboni}, {Casu}, {Concu}, {Corongiu}, {Deiana}, {Egron},
  {Fara}, {Gaudiomonte}, {Gusai}, {Ladu}, {Loru}, {Leurini}, {Marongiu},
  {Melis}, {Melis}, {Migoni}, {Milia}, {Navarrini}, {Orlati}, {Ortu}, {Palmas},
  {Pellizzoni}, {Perrodin}, {Pisanu}, {Poppi}, {Righini}, {Saba}, {Serra},
  {Serrau}, {Stagni}, {Surcis}, {Vacca}, {Vargiu}, {Hunt}, {Jin}, {Klose},
  {Kouveliotou}, {Mazzali}, {M{\o}ller}, {Nava}, {Piran}, {Selsing}, {Vergani},
  {Wiersema}, {Toma}, {Higgins}, {Mundell}, {di Serego Alighieri}, {G{\'o}tz},
  {Gao}, {Gomboc}, {Kaper}, {Kobayashi}, {Kopac}, {Mao}, {Starling}, {Steele},
  {van der Horst}, {GRAWITA: GRAvitational Wave Inaf TeAm}, {Acero}, {Atwood},
  {Baldini}, {Barbiellini}, {Bastieri}, {Berenji}, {Bellazzini}, {Bissaldi},
  {Blandford}, {Bloom}, {Bonino}, {Bottacini}, {Bregeon}, {Buehler}, {Buson},
  {Cameron}, {Caputo}, {Caraveo}, {Cavazzuti}, {Chekhtman}, {Cheung}, {Chiang},
  {Ciprini}, {Cohen-Tanugi}, {Cominsky}, {Costantin}, {Cuoco}, {D'Ammando}, {de
  Palma}, {Digel}, {Di Lalla}, {Di Mauro}, {Di Venere}, {Dubois}, {Fegan},
  {Focke}, {Franckowiak}, {Fukazawa}, {Funk}, {Fusco}, {Gargano}, {Gasparrini},
  {Giglietto}, {Giordano}, {Giroletti}, {Glanzman}, {Green}, {Grondin},
  {Guillemot}, {Guiriec}, {Harding}, {Horan}, {J{\'o}hannesson}, {Kamae},
  {Kensei}, {Kuss}, {La Mura}, {Latronico}, {Lemoine-Goumard}, {Longo},
  {Loparco}, {Lovellette}, {Lubrano}, {Magill}, {Maldera}, {Manfreda},
  {Mazziotta}, {McEnery}, {Meyer}, {Michelson}, {Mirabal}, {Monzani},
  {Moretti}, {Morselli}, {Moskalenko}, {Negro}, {Nuss}, {Ojha}, {Omodei},
  {Orienti}, {Orlando}, {Palatiello}, {Paliya}, {Paneque}, {Pesce-Rollins},
  {Piron}, {Porter}, {Principe}, {Rain{\`o}}, {Rando}, {Razzano}, {Razzaque},
  {Reimer}, {Reimer}, {Reposeur}, {Rochester}, {Saz Parkinson}, {Sgr{\`o}},
  {Siskind}, {Spada}, {Spandre}, {Suson}, {Takahashi}, {Tanaka}, {Thayer},
  {Thayer}, {Thompson}, {Tibaldo}, {Torres}, {Torresi}, {Troja}, {Venters},
  {Vianello}, {Zaharijas}, {Fermi Large Area Telescope Collaboration},
  {Allison}, {Bannister}, {Dobie}, {Kaplan}, {Lenc}, {Lynch}, {Murphy},
  {Sadler}, {Australia Telescope Compact Array}, {Hotan}, {James}, {Oslowski},
  {Raja}, {Shannon}, {Whiting}, {Australian SKA Pathfinder}, {Arcavi},
  {Howell}, {McCully}, {Hosseinzadeh}, {Hiramatsu}, {Poznanski}, {Barnes},
  {Zaltzman}, {Vasylyev}, {Maoz}, {Las Cumbres Observatory Group}, {Cooke},
  {Bailes}, {Wolf}, {Deller}, {Lidman}, {Wang}, {Gendre}, {Andreoni}, {Ackley},
  {Pritchard}, {Bessell}, {Chang}, {M{\"o}ller}, {Onken}, {Scalzo},
  {Ridden-Harper}, {Sharp}, {Tucker}, {Farrell}, {Elmer}, {Johnston},
  {Venkatraman Krishnan}, {Keane}, {Green}, {Jameson}, {Hu}, {Ma}, {Sun}, {Wu},
  {Wang}, {Shang}, {Hu}, {Ashley}, {Yuan}, {Li}, {Tao}, {Zhu}, {Zhang},
  {Suntzeff}, {Zhou}, {Yang}, {Orange}, {Morris}, {Cucchiara}, {Giblin},
  {Klotz}, {Staff}, {Thierry}, {Schmidt}, {OzGrav}, {(Deeper}, {Wider},
  {program}, {AST3}, {CAASTRO Collaborations}, {Tanvir}, {Levan}, {Cano}, {de
  Ugarte-Postigo}, {Gonz{\'a}lez-Fern{\'a}ndez}, {Greiner}, {Hjorth}, {Irwin},
  {Kr{\"u}hler}, {Mandel}, {Milvang-Jensen}, {O'Brien}, {Rol}, {Rosetti},
  {Rosswog}, {Rowlinson}, {Steeghs}, {Th{\"o}ne}, {Ulaczyk}, {Watson}, {Bruun},
  {Cutter}, {Figuera Jaimes}, {Fujii}, {Fruchter}, {Gompertz}, {Jakobsson},
  {Hodosan}, {J{\`e}rgensen}, {Kangas}, {Kann}, {Rabus}, {Schr{\o}der},
  {Stanway}, {Wijers}, {VINROUGE Collaboration}, {Lipunov}, {Gorbovskoy},
  {Kornilov}, {Tyurina}, {Balanutsa}, {Kuznetsov}, {Vlasenko}, {Podesta},
  {Lopez}, {Podesta}, {Levato}, {Saffe}, {Mallamaci}, {Budnev}, {Gress},
  {Kuvshinov}, {Gorbunov}, {Vladimirov}, {Zimnukhov}, {Gabovich}, {Yurkov},
  {Sergienko}, {Rebolo}, {Serra-Ricart}, {Tlatov}, {Ishmuhametova}, {MASTER
  Collaboration}, {Abe}, {Aoki}, {Aoki}, {Asakura}, {Baar}, {Barway}, {Bond},
  {Doi}, {Finet}, {Fujiyoshi}, {Furusawa}, {Honda}, {Itoh}, {Kanda},
  {Kawabata}, {Kawabata}, {Kim}, {Koshida}, {Kuroda}, {Lee}, {Liu},
  {Matsubayashi}, {Miyazaki}, {Morihana}, {Morokuma}, {Motohara}, {Murata},
  {Nagai}, {Nagashima}, {Nagayama}, {Nakaoka}, {Nakata}, {Ohsawa}, {Ohshima},
  {Ohta}, {Okita}, {Saito}, {Saito}, {Sako}, {Sekiguchi}, {Sumi}, {Tajitsu},
  {Takahashi}, {Takayama}, {Tamura}, {Tanaka}, {Tanaka}, {Terai}, {Tominaga},
  {Tristram}, {Uemura}, {Utsumi}, {Yamaguchi}, {Yasuda}, {Yoshida}, {Zenko},
  {J-GEM}, {Adams}, {Anupama}, {Bally}, {Barway}, {Bellm}, {Blagorodnova},
  {Cannella}, {Chandra}, {Chatterjee}, {Clarke}, {Cobb}, {Cook}, {Copperwheat},
  {De}, {Emery}, {Feindt}, {Foster}, {Fox}, {Frail}, {Fremling}, {Frohmaier},
  {Garcia}, {Ghosh}, {Giacintucci}, {Goobar}, {Gottlieb}, {Grefenstette},
  {Hallinan}, {Harrison}, {Heida}, {Helou}, {Ho}, {Horesh}, {Hotokezaka}, {Ip},
  {Itoh}, {Jacobs}, {Jencson}, {Kasen}, {Kasliwal}, {Kassim}, {Kim}, {Kiran},
  {Kuin}, {Kulkarni}, {Kupfer}, {Lau}, {Madsen}, {Mazzali}, {Miller},
  {Miyasaka}, {Mooley}, {Myers}, {Nakar}, {Ngeow}, {Nugent}, {Ofek},
  {Palliyaguru}, {Pavana}, {Perley}, {Peters}, {Pike}, {Piran}, {Qi}, {Quimby},
  {Rana}, {Rosswog}, {Rusu}, {Sadler}, {Van Sistine}, {Sollerman}, {Xu}, {Yan},
  {Yatsu}, {Yu}, {Zhang}, {Zhao}, {GROWTH}, {JAGWAR}, {Caltech-NRAO},
  {TTU-NRAO}, {NuSTAR Collaborations}, {Chambers}, {Huber}, {Schultz},
  {Bulger}, {Flewelling}, {Magnier}, {Lowe}, {Wainscoat}, {Waters}, {Willman},
  {Pan-STARRS}, {Ebisawa}, {Hanyu}, {Harita}, {Hashimoto}, {Hidaka}, {Hori},
  {Ishikawa}, {Isobe}, {Iwakiri}, {Kawai}, {Kawai}, {Kawamuro}, {Kawase},
  {Kitaoka}, {Makishima}, {Matsuoka}, {Mihara}, {Morita}, {Morita}, {Nakahira},
  {Nakajima}, {Nakamura}, {Negoro}, {Oda}, {Sakamaki}, {Sasaki}, {Serino},
  {Shidatsu}, {Shimomukai}, {Sugawara}, {Sugita}, {Sugizaki}, {Tachibana},
  {Takao}, {Tanimoto}, {Tomida}, {Tsuboi}, {Tsunemi}, {Ueda}, {Ueno}, {Yamada},
  {Yamaoka}, {Yamauchi}, {Yatabe}, {Yoneyama}, {Yoshii}, {MAXI Team}, {Coward},
  {Crisp}, {Macpherson}, {Andreoni}, {Laugier}, {Noysena}, {Klotz}, {Gendre},
  {Thierry}, {Turpin}, {Consortium}, {Im}, {Choi}, {Kim}, {Yoon}, {Lim}, {Lee},
  {Lee}, {Kim}, {Ko}, {Joe}, {Kwon}, {Kim}, {Lim}, {Choi}, {KU Collaboration},
  {Fynbo}, {Malesani}, {Xu}, {Optical Telescope}, {Smartt}, {Jerkstrand},
  {Kankare}, {Sim}, {Fraser}, {Inserra}, {Maguire}, {Leloudas}, {Magee},
  {Shingles}, {Smith}, {Young}, {Kotak}, {Gal-Yam}, {Lyman}, {Homan},
  {Agliozzo}, {Anderson}, {Angus}, {Ashall}, {Barbarino}, {Bauer}, {Berton},
  {Botticella}, {Bulla}, {Cannizzaro}, {Cartier}, {Cikota}, {Clark}, {De Cia},
  {Della Valle}, {Dennefeld}, {Dessart}, {Dimitriadis}, {Elias-Rosa}, {Firth},
  {Fl{\"o}rs}, {Frohmaier}, {Galbany}, {Gonz{\'a}lez-Gait{\'a}n}, {Gromadzki},
  {Guti{\'e}rrez}, {Hamanowicz}, {Harmanen}, {Heintz}, {Hernandez}, {Hodgkin},
  {Hook}, {Izzo}, {James}, {Jonker}, {Kerzendorf}, {Kostrzewa-Rutkowska},
  {Kromer}, {Kuncarayakti}, {Lawrence}, {Manulis}, {Mattila}, {McBrien},
  {M{\"u}ller}, {Nordin}, {O'Neill}, {Onori}, {Palmerio}, {Pastorello},
  {Patat}, {Pignata}, {Podsiadlowski}, {Razza}, {Reynolds}, {Roy}, {Ruiter},
  {Rybicki}, {Salmon}, {Pumo}, {Prentice}, {Seitenzahl}, {Smith}, {Sollerman},
  {Sullivan}, {Szegedi}, {Taddia}, {Taubenberger}, {Terreran}, {Van Soelen},
  {Vos}, {Walton}, {Wright}, {Wyrzykowski}, {Yaron}, {pre=''(''>ePESSTO},
  {Chen}, {Kr{\"u}hler}, {Schady}, {Wiseman}, {Greiner}, {Rau}, {Schweyer},
  {Klose}, {Nicuesa Guelbenzu}, {GROND}, {Palliyaguru}, {Tech University},
  {Shara}, {Williams}, {Vaisanen}, {Potter}, {Romero Colmenero}, {Crawford},
  {Buckley}, {Mao}, {SALT Group}, {D{\'\i}az}, {Macri}, {Garc{\'\i}a Lambas},
  {Mendes de Oliveira}, {Nilo Castell{\'o}n}, {Ribeiro}, {S{\'a}nchez},
  {Schoenell}, {Abramo}, {Akras}, {Alcaniz}, {Artola}, {Beroiz}, {Bonoli},
  {Cabral}, {Camuccio}, {Chavushyan}, {Coelho}, {Colazo}, {Costa-Duarte},
  {Cuevas Larenas}, {Dom{\'\i}nguez Romero}, {Dultzin}, {Fern{\'a}ndez},
  {Garc{\'\i}a}, {Girardini}, {Gon{\c{c}}alves}, {Gon{\c{c}}alves}, {Gurovich},
  {Jim{\'e}nez-Teja}, {Kanaan}, {Lares}, {Lopes de Oliveira}, {L{\'o}pez-Cruz},
  {Melia}, {Molino}, {Padilla}, {Pe{\~n}uela}, {Placco}, {Qui{\~n}ones},
  {Ram{\'\i}rez Rivera}, {Renzi}, {Riguccini}, {R{\'\i}os-L{\'o}pez},
  {Rodriguez}, {Sampedro}, {Schneiter}, {Sodr{\'e}}, {Starck}, {Torres-Flores},
  {Tornatore}, {Zadro{\.z}ny}, {Castillo}, {TOROS: Transient Robotic
  Observatory of South Collaboration}, {Castro-Tirado}, {Tello}, {Hu}, {Zhang},
  {Cunniffe}, {Castell{\'o}n}, {Hiriart}, {Caballero-Garc{\'\i}a},
  {Jel{\'\i}nek}, {Kub{\'a}nek}, {P{\'e}rez del Pulgar}, {Park}, {Jeong},
  {Castro Cer{\'o}n}, {Pandey}, {Yock}, {Querel}, {Fan}, {Wang}, {BOOTES
  Collaboration}, {Beardsley}, {Brown}, {Crosse}, {Emrich}, {Franzen},
  {Gaensler}, {Horsley}, {Johnston-Hollitt}, {Kenney}, {Morales}, {Pallot},
  {Sokolowski}, {Steele}, {Tingay}, {Trott}, {Walker}, {Wayth}, {Williams},
  {Wu}, {Murchison Widefield Array}, {Yoshida}, {Sakamoto}, {Kawakubo},
  {Yamaoka}, {Takahashi}, {Asaoka}, {Ozawa}, {Torii}, {Shimizu}, {Tamura},
  {Ishizaki}, {Cherry}, {Ricciarini}, {Penacchioni}, {Marrocchesi}, {CALET
  Collaboration}, {Pozanenko}, {Volnova}, {Mazaeva}, {Minaev}, {Krugov},
  {Kusakin}, {Reva}, {Moskvitin}, {Rumyantsev}, {Inasaridze}, {Klunko},
  {Tungalag}, {Schmalz}, {Burhonov}, {IKI-GW Follow-up Collaboration},
  {Abdalla}, {Abramowski}, {Aharonian}, {Ait Benkhali}, {Ang{\"u}ner},
  {Arakawa}, {Arrieta}, {Aubert}, {Backes}, {Balzer}, {Barnard}, {Becherini},
  {Becker Tjus}, {Berge}, {Bernhard}, {Bernl{\"o}hr}, {Blackwell},
  {B{\"o}ttcher}, {Boisson}, {Bolmont}, {Bonnefoy}, {Bordas}, {Bregeon},
  {Brun}, {Brun}, {Bryan}, {B{\"u}chele}, {Bulik}, {Capasso}, {Caroff},
  {Carosi}, {Casanova}, {Cerruti}, {Chakraborty}, {Chaves}, {Chen},
  {Chevalier}, {Colafrancesco}, {Condon}, {Conrad}, {Davids}, {Decock}, {Deil},
  {Devin}, {deWilt}, {Dirson}, {Djannati-Ata{\"\i}}, {Donath}, {O'C. Drury},
  {Dutson}, {Dyks}, {Edwards}, {Egberts}, {Emery}, {Ernenwein}, {Eschbach},
  {Farnier}, {Fegan}, {Fernandes}, {Fiasson}, {Fontaine}, {Funk},
  {F{\"u}ssling}, {Gabici}, {Gallant}, {Garrigoux}, {Gat{\'e}}, {Giavitto},
  {Giebels}, {Glawion}, {Glicenstein}, {Gottschall}, {Grondin}, {Hahn},
  {Haupt}, {Hawkes}, {Heinzelmann}, {Henri}, {Hermann}, {Hinton}, {Hofmann},
  {Hoischen}, {Holch}, {Holler}, {Horns}, {Ivascenko}, {Iwasaki},
  {Jacholkowska}, {Jamrozy}, {Jankowsky}, {Jankowsky}, {Jingo}, {Jouvin},
  {Jung-Richardt}, {Kastendieck}, {Katarzy{\'n}ski}, {Katsuragawa},
  {Kerszberg}, {Khangulyan}, {Kh{\'e}lifi}, {King}, {Klepser}, {Klochkov},
  {Klu{\'z}niak}, {Komin}, {Kosack}, {Krakau}, {Kraus}, {Kr{\"u}ger}, {Laffon},
  {Lamanna}, {Lau}, {Lees}, {Lefaucheur}, {Lemi{\`e}re}, {Lemoine-Goumard},
  {Lenain}, {Leser}, {Lohse}, {Lorentz}, {Liu}, {Lypova}, {Malyshev},
  {Marandon}, {Marcowith}, {Mariaud}, {Marx}, {Maurin}, {Maxted}, {Mayer},
  {Meintjes}, {Meyer}, {Mitchell}, {Moderski}, {Mohamed}, {Mohrmann},
  {Mor{\r{a}}}, {Moulin}, {Murach}, {Nakashima}, {de Naurois}, {Ndiyavala},
  {Niederwanger}, {Niemiec}, {Oakes}, {O'Brien}, {Odaka}, {Ohm}, {Ostrowski},
  {Oya}, {Padovani}, {Panter}, {Parsons}, {Pekeur}, {Pelletier}, {Perennes},
  {Petrucci}, {Peyaud}, {Piel}, {Pita}, {Poireau}, {Poon}, {Prokhorov},
  {Prokoph}, {P{\"u}hlhofer}, {Punch}, {Quirrenbach}, {Raab}, {Rauth},
  {Reimer}, {Reimer}, {Renaud}, {de los Reyes}, {Rieger}, {Rinchiuso},
  {Romoli}, {Rowell}, {Rudak}, {Rulten}, {Sahakian}, {Saito}, {Sanchez},
  {Santangelo}, {Sasaki}, {Schlickeiser}, {Sch{\"u}ssler}, {Schulz},
  {Schwanke}, {Schwemmer}, {Seglar-Arroyo}, {Settimo}, {Seyffert}, {Shafi},
  {Shilon}, {Shiningayamwe}, {Simoni}, {Sol}, {Spanier}, {Spir-Jacob},
  {Stawarz}, {Steenkamp}, {Stegmann}, {Steppa}, {Sushch}, {Takahashi},
  {Tavernet}, {Tavernier}, {Taylor}, {Terrier}, {Tibaldo}, {Tiziani},
  {Tluczykont}, {Trichard}, {Tsirou}, {Tsuji}, {Tuffs}, {Uchiyama}, {van der
  Walt}, {van Eldik}, {van Rensburg}, {van Soelen}, {Vasileiadis}, {Veh},
  {Venter}, {Viana}, {Vincent}, {Vink}, {Voisin}, {V{\"o}lk}, {Vuillaume},
  {Wadiasingh}, {Wagner}, {Wagner}, {Wagner}, {White}, {Wierzcholska},
  {Willmann}, {W{\"o}rnlein}, {Wouters}, {Yang}, {Zaborov}, {Zacharias},
  {Zanin}, {Zdziarski}, {Zech}, {Zefi}, {Ziegler}, {Zorn}, {{\.Z}ywucka},
  {H.~E.~S.~S. Collaboration}, {Fender}, {Broderick}, {Rowlinson}, {Wijers},
  {Stewart}, {ter Veen}, {Shulevski}, {LOFAR Collaboration}, {Kavic},
  {Simonetti}, {League}, {Tsai}, {Obenberger}, {Nathaniel}, {Taylor}, {Dowell},
  {Liebling}, {Estes}, {Lippert}, {Sharma}, {Vincent}, {Farella}, {Wavelength
  Array}, {Abeysekara}, {Albert}, {Alfaro}, {Alvarez}, {Arceo},
  {Arteaga-Vel{\'a}zquez}, {Avila Rojas}, {Ayala Solares}, {Barber}, {Becerra
  Gonzalez}, {Becerril}, {Belmont-Moreno}, {BenZvi}, {Berley}, {Bernal},
  {Braun}, {Brisbois}, {Caballero-Mora}, {Capistr{\'a}n}, {Carrami{\~n}ana},
  {Casanova}, {Castillo}, {Cotti}, {Cotzomi}, {Couti{\~n}o de Le{\'o}n}, {De
  Le{\'o}n}, {De la Fuente}, {Diaz Hernandez}, {Dichiara}, {Dingus},
  {DuVernois}, {D{\'\i}az-V{\'e}lez}, {Ellsworth}, {Engel},
  {Enr{\'\i}quez-Rivera}, {Fiorino}, {Fleischhack}, {Fraija},
  {Garc{\'\i}a-Gonz{\'a}lez}, {Garfias}, {Gerhardt}, {Gonz{\~o}lez Mu{\~n}oz},
  {Gonz{\'a}lez}, {Goodman}, {Hampel-Arias}, {Harding}, {Hernandez},
  {Hernandez-Almada}, {Hona}, {H{\"u}ntemeyer}, {Iriarte}, {Jardin-Blicq},
  {Joshi}, {Kaufmann}, {Kieda}, {Lara}, {Lauer}, {Lennarz}, {Le{\'o}n Vargas},
  {Linnemann}, {Longinotti}, {Raya}, {Luna-Garc{\'\i}a}, {L{\'o}pez-Coto},
  {Malone}, {Marinelli}, {Martinez}, {Martinez-Castellanos},
  {Mart{\'\i}nez-Castro}, {Mart{\'\i}nez-Huerta}, {Matthews},
  {Miranda-Romagnoli}, {Moreno}, {Mostaf{\'a}}, {Nellen}, {Newbold}, {Nisa},
  {Noriega-Papaqui}, {Pelayo}, {Pretz}, {P{\'e}rez-P{\'e}rez}, {Ren}, {Rho},
  {Rivi{\`e}re}, {Rosa-Gonz{\'a}lez}, {Rosenberg}, {Ruiz-Velasco}, {Salazar},
  {Salesa Greus}, {Sandoval}, {Schneider}, {Schoorlemmer}, {Sinnis}, {Smith},
  {Springer}, {Surajbali}, {Tibolla}, {Tollefson}, {Torres}, {Ukwatta},
  {Weisgarber}, {Westerhoff}, {Wisher}, {Wood}, {Yapici}, {Yodh}, {Younk},
  {Zhou}, {{\'A}lvarez}, {HAWC Collaboration}, {Aab}, {Abreu}, {Aglietta},
  {Albuquerque}, {Albury}, {Allekotte}, {Almela}, {Alvarez Castillo},
  {Alvarez-Mu{\~n}iz}, {Anastasi}, {Anchordoqui}, {Andrada}, {Andringa},
  {Aramo}, {Arsene}, {Asorey}, {Assis}, {Avila}, {Badescu}, {Balaceanu},
  {Barbato}, {Barreira Luz}, {Becker}, {Bellido}, {Berat}, {Bertaina},
  {Bertou}, {Biermann}, {Biteau}, {Blaess}, {Blanco}, {Blazek}, {Bleve},
  {Boh{\'a}{\v{c}}ov{\'a}}, {Bonifazi}, {Borodai}, {Botti}, {Brack}, {Brancus},
  {Bretz}, {Bridgeman}, {Briechle}, {Buchholz}, {Bueno}, {Buitink}, {Buscemi},
  {Caballero-Mora}, {Caccianiga}, {Cancio}, {Canfora}, {Caruso}, {Castellina},
  {Catalani}, {Cataldi}, {Cazon}, {Chavez}, {Chinellato}, {Chudoba}, {Clay},
  {Cobos Cerutti}, {Colalillo}, {Coleman}, {Collica}, {Coluccia},
  {Concei{\c{c}}{\~a}o}, {Consolati}, {Contreras}, {Cooper}, {Coutu},
  {Covault}, {Cronin}, {D'Amico}, {Daniel}, {Dasso}, {Daumiller}, {Dawson},
  {Day}, {de Almeida}, {de Jong}, {De Mauro}, {de Mello Neto}, {De Mitri}, {de
  Oliveira}, {de Souza}, {Debatin}, {Deligny}, {D{\'\i}az Castro}, {Diogo},
  {Dobrigkeit}, {D'Olivo}, {Dorosti}, {Dos Anjos}, {Dova}, {Dundovic}, {Ebr},
  {Engel}, {Erdmann}, {Erfani}, {Escobar}, {Espadanal}, {Etchegoyen}, {Falcke},
  {Farmer}, {Farrar}, {Fauth}, {Fazzini}, {Feldbusch}, {Fenu}, {Fick},
  {Figueira}, {Filip{\v{c}}i{\v{c}}}, {Freire}, {Fujii}, {Fuster},
  {Ga{\"\i}or}, {Garc{\'\i}a}, {Gat{\'e}}, {Gemmeke}, {Gherghel-Lascu}, {Ghia},
  {Giaccari}, {Giammarchi}, {Giller}, {G{\l}as}, {Glaser}, {Golup}, {G{\'o}mez
  Berisso}, {G{\'o}mez Vitale}, {Gonz{\'a}lez}, {Gorgi}, {Gottowik}, {Grillo},
  {Grubb}, {Guarino}, {Guedes}, {Halliday}, {Hampel}, {Hansen}, {Harari},
  {Harrison}, {Harvey}, {Haungs}, {Hebbeker}, {Heck}, {Heimann}, {Herve},
  {Hill}, {Hojvat}, {Holt}, {Homola}, {H{\"o}randel}, {Horvath},
  {Hrabovsk{\'y}}, {Huege}, {Hulsman}, {Insolia}, {Isar}, {Jandt}, {Johnsen},
  {Josebachuili}, {Jurysek}, {K{\"a}{\"a}p{\"a}}, {Kampert}, {Keilhauer},
  {Kemmerich}, {Kemp}, {Kieckhafer}, {Klages}, {Kleifges}, {Kleinfeller},
  {Krause}, {Krohm}, {Kuempel}, {Kukec Mezek}, {Kunka}, {Kuotb Awad}, {Lago},
  {LaHurd}, {Lang}, {Lauscher}, {Legumina}, {Leigui de Oliveira},
  {Letessier-Selvon}, {Lhenry-Yvon}, {Link}, {Lo Presti}, {Lopes}, {L{\'o}pez},
  {L{\'o}pez Casado}, {Lorek}, {Luce}, {Lucero}, {Malacari}, {Mallamaci},
  {Mandat}, {Mantsch}, {Mariazzi}, {Maris}, {Marsella}, {Martello}, {Martinez},
  {Mart{\'\i}nez Bravo}, {Mas{\'\i}as Meza}, {Mathes}, {Mathys}, {Matthews},
  {Matthiae}, {Mayotte}, {Mazur}, {Medina}, {Medina-Tanco}, {Melo},
  {Menshikov}, {Merenda}, {Michal}, {Micheletti}, {Middendorf}, {Miramonti},
  {Mitrica}, {Mockler}, {Mollerach}, {Montanet}, {Morello}, {Morlino},
  {M{\"u}ller}, {M{\"u}ller}, {Muller}, {M{\"u}ller}, {Mussa}, {Naranjo},
  {Nguyen}, {Niculescu-Oglinzanu}, {Niechciol}, {Niemietz}, {Niggemann},
  {Nitz}, {Nosek}, {Novotny}, {No{\v{z}}ka}, {N{\'u}{\~n}ez}, {Oikonomou},
  {Olinto}, {Palatka}, {Pallotta}, {Papenbreer}, {Parente}, {Parra}, {Paul},
  {Pech}, {Pedreira}, {P{\c{e}}kala}, {Pe{\~n}a-Rodriguez}, {Pereira},
  {Perlin}, {Perrone}, {Peters}, {Petrera}, {Phuntsok}, {Pierog}, {Pimenta},
  {Pirronello}, {Platino}, {Plum}, {Poh}, {Porowski}, {Prado}, {Privitera},
  {Prouza}, {Quel}, {Querchfeld}, {Quinn}, {Ramos-Pollan}, {Rautenberg},
  {Ravignani}, {Ridky}, {Riehn}, {Risse}, {Ristori}, {Rizi}, {Rodrigues de
  Carvalho}, {Rodriguez Fernandez}, {Rodriguez Rojo}, {Roncoroni}, {Roth},
  {Roulet}, {Rovero}, {Ruehl}, {Saffi}, {Saftoiu}, {Salamida}, {Salazar},
  {Saleh}, {Salina}, {S{\'a}nchez}, {Sanchez-Lucas}, {Santos}, {Santos},
  {Sarazin}, {Sarmento}, {Sarmiento-Cano}, {Sato}, {Schauer}, {Scherini},
  {Schieler}, {Schimp}, {Schmidt}, {Scholten}, {Schov{\'a}nek}, {Schr{\"o}der},
  {Schr{\"o}der}, {Schulz}, {Schumacher}, {Sciutto}, {Segreto}, {Shadkam},
  {Shellard}, {Sigl}, {Silli}, {{\v{S}}m{\'\i}da}, {Snow}, {Sommers},
  {Sonntag}, {Soriano}, {Squartini}, {Stanca}, {Stani{\v{c}}}, {Stasielak},
  {Stassi}, {Stolpovskiy}, {Strafella}, {Streich}, {Suarez},
  {Suarez-Dur{\'a}n}, {Sudholz}, {Suomij{\"a}rvi}, {Supanitsky},
  {{\v{S}}up{\'\i}k}, {Swain}, {Szadkowski}, {Taboada}, {Taborda},
  {Timmermans}, {Todero Peixoto}, {Tomankova}, {Tom{\'e}}, {Torralba Elipe},
  {Travnicek}, {Trini}, {Tueros}, {Ulrich}, {Unger}, {Urban}, {Vald{\'e}s
  Galicia}, {Vali{\~n}o}, {Valore}, {van Aar}, {van Bodegom}, {van den Berg},
  {van Vliet}, {Varela}, {Vargas C{\'a}rdenas}, {V{\'a}zquez}, {Veberi{\v{c}}},
  {Ventura}, {Vergara Quispe}, {Verzi}, {Vicha}, {Villase{\~n}or}, {Vorobiov},
  {Wahlberg}, {Wainberg}, {Walz}, {Watson}, {Weber}, {Weindl}, {Wiede{\'n}ski},
  {Wiencke}, {Wilczy{\'n}ski}, {Wirtz}, {Wittkowski}, {Wundheiler}, {Yang},
  {Yushkov}, {Zas}, {Zavrtanik}, {Zavrtanik}, {Zepeda}, {Zimmermann},
  {Ziolkowski}, {Zong}, {Zuccarello}, {Pierre Auger Collaboration}, {Kim},
  {Schulze}, {Bauer}, {Corral-Santana}, {de Gregorio-Monsalvo},
  {Gonz{\'a}lez-L{\'o}pez}, {Hartmann}, {Ishwara-Chandra}, {Mart{\'\i}n},
  {Mehner}, {Misra}, {Micha{\l}owski}, {Resmi}, {ALMA Collaboration}, {Paragi},
  {Agudo}, {An}, {Beswick}, {Casadio}, {Frey}, {Jonker}, {Kettenis}, {Marcote},
  {Moldon}, {Szomoru}, {van Langevelde}, {Yang}, {Euro VLBI Team}, {Cwiek},
  {Cwiok}, {Czyrkowski}, {Dabrowski}, {Kasprowicz}, {Mankiewicz}, {Nawrocki},
  {Opiela}, {Piotrowski}, {Wrochna}, {Zaremba}, {{\.Z}arnecki}, {Pi of Sky
  Collaboration}, {Haggard}, {Nynka}, {Ruan}, {Chandra Team at McGill
  University}, {Bland}, {Booler}, {Devillepoix}, {de Gois}, {Hancock}, {Howie},
  {Paxman}, {Sansom}, {Towner}, {Desert Fireball Network}, {Tonry}, {Coughlin},
  {Stubbs}, {Denneau}, {Heinze}, {Stalder}, {Weiland}, {ATLAS}, {Eatough},
  {Kramer}, {Kraus}, {Time Resolution Universe Survey}, {Troja}, {Piro},
  {Becerra Gonz{\'a}lez}, {Butler}, {Fox}, {Khandrika}, {Kutyrev}, {Lee},
  {Ricci}, {Ryan}, {S{\'a}nchez-Ram{\'\i}rez}, {Veilleux}, {Watson},
  {Wieringa}, {Burgess}, {van Eerten}, {Fontes}, {Fryer}, {Korobkin},
  {Wollaeger}, {RIMAS}, {RATIR}, {Camilo}, {Foley}, {Goedhart}, {Makhathini},
  {Oozeer}, {Smirnov}, {Fender}, {Woudt}, \& {South
  Africa/MeerKAT}}]{AbbottGW17}
{Abbott}, B.~P., {Abbott}, R., {Abbott}, T.~D., {et~al.} 2017, \apjl, 848, L12,
  \dodoi{10.3847/2041-8213/aa91c9}

\bibitem[{{Abbott} {et~al.}(2020{\natexlab{a}}){Abbott}, {Abbott}, {Abbott},
  {Abraham}, {Acernese}, {Ackley}, {Adams}, {Adhikari}, {Adya}, {Affeldt},
  {Agathos}, {Agatsuma}, {Aggarwal}, {Aguiar}, {Aiello}, {Ain}, {Ajith},
  {Allen}, {Allocca}, {Aloy}, {Altin}, {Amato}, {Anand}, {Ananyeva},
  {Anderson}, {Anderson}, {Angelova}, {Antier}, {Appert}, {Arai}, {Araya},
  {Areeda}, {Ar{\`e}ne}, {Arnaud}, {Aronson}, {Arun}, {Ascenzi}, {Ashton},
  {Aston}, {Astone}, {Aubin}, {Aufmuth}, {AultONeal}, {Austin}, {Avendano},
  {Avila-Alvarez}, {Babak}, {Bacon}, {Badaracco}, {Bader}, {Bae}, {Baird},
  {Baker}, {Baldaccini}, {Ballardin}, {Ballmer}, {Bals}, {Banagiri},
  {Barayoga}, {Barbieri}, {Barclay}, {Barish}, {Barker}, {Barkett}, {Barnum},
  {Barone}, {Barr}, {Barsotti}, {Barsuglia}, {Barta}, {Bartlett}, {Bartos},
  {Bassiri}, {Basti}, {Bawaj}, {Bayley}, {Baylor}, {Bazzan}, {B{\'e}csy},
  {Bejger}, {Belahcene}, {Bell}, {Beniwal}, {Benjamin}, {Berger}, {Bergmann},
  {Bernuzzi}, {Berry}, {Bersanetti}, {Bertolini}, {Betzwieser}, {Bhandare},
  {Bidler}, {Biggs}, {Bilenko}, {Bilgili}, {Billingsley}, {Birney},
  {Birnholtz}, {Biscans}, {Bischi}, {Biscoveanu}, {Bisht}, {Bitossi},
  {Bizouard}, {Blackburn}, {Blackman}, {Blair}, {Blair}, {Blair}, {Bloemen},
  {Bobba}, {Bode}, {Boer}, {Boetzel}, {Bogaert}, {Bondu}, {Bonnand}, {Booker},
  {Boom}, {Bork}, {Boschi}, {Bose}, {Bossilkov}, {Bosveld}, {Bouffanais},
  {Bozzi}, {Bradaschia}, {Brady}, {Bramley}, {Branchesi}, {Brau}, {Breschi},
  {Briant}, {Briggs}, {Brighenti}, {Brillet}, {Brinkmann}, {Brockill},
  {Brooks}, {Brooks}, {Brown}, {Brunett}, {Buikema}, {Bulik}, {Bulten},
  {Buonanno}, {Buskulic}, {Buy}, {Byer}, {Cabero}, {Cadonati}, {Cagnoli},
  {Cahillane}, {Calder{\'o}n Bustillo}, {Callister}, {Calloni}, {Camp},
  {Campbell}, {Canepa}, {Cannon}, {Cao}, {Cao}, {Carapella}, {Carbognani},
  {Caride}, {Carney}, {Carullo}, {Casanueva Diaz}, {Casentini}, {Caudill},
  {Cavagli{\`a}}, {Cavalier}, {Cavalieri}, {Cella}, {Cerd{\'a}-Dur{\'a}n},
  {Cesarini}, {Chaibi}, {Chakravarti}, {Chamberlin}, {Chan}, {Chao},
  {Charlton}, {Chase}, {Chassande-Mottin}, {Chatterjee}, {Chaturvedi},
  {Chatziioannou}, {Cheeseboro}, {Chen}, {Chen}, {Chen}, {Cheng}, {Cheong},
  {Chia}, {Chiadini}, {Chincarini}, {Chiummo}, {Cho}, {Cho}, {Cho},
  {Christensen}, {Chu}, {Chua}, {Chung}, {Chung}, {Ciani}, {Cie{\'s}lar},
  {Ciobanu}, {Ciolfi}, {Cipriano}, {Cirone}, {Clara}, {Clark}, {Clearwater},
  {Cleva}, {Coccia}, {Cohadon}, {Cohen}, {Colleoni}, {Collette}, {Collins},
  {Colpi}, {Cominsky}, {Constancio}, {Conti}, {Cooper}, {Corban}, {Corbitt},
  {Cordero-Carri{\'o}n}, {Corezzi}, {Corley}, {Cornish}, {Corre}, {Corsi},
  {Cortese}, {Costa}, {Cotesta}, {Coughlin}, {Coughlin}, {Coulon},
  {Countryman}, {Couvares}, {Covas}, {Cowan}, {Coward}, {Cowart}, {Coyne},
  {Coyne}, {Creighton}, {Creighton}, {Cripe}, {Croquette}, {Crowder}, {Cullen},
  {Cumming}, {Cunningham}, {Cuoco}, {Dal Canton}, {D{\'a}lya}, {D'Angelo},
  {Danilishin}, {D'Antonio}, {Danzmann}, {Dasgupta}, {Da Silva Costa},
  {Datrier}, {Dattilo}, {Dave}, {Davier}, {Davis}, {Daw}, {DeBra},
  {Deenadayalan}, {Degallaix}, {De Laurentis}, {Del{\'e}glise}, {De Lillo},
  {Del Pozzo}, {DeMarchi}, {Demos}, {Dent}, {De Pietri}, {De Rosa}, {De Rossi},
  {DeSalvo}, {de Varona}, {Dhurandhar}, {D{\'\i}az}, {Dietrich}, {Di Fiore},
  {DiFronzo}, {Di Giorgio}, {Di Giovanni}, {Di Giovanni}, {Di Girolamo}, {Di
  Lieto}, {Ding}, {Di Pace}, {Di Palma}, {Di Renzo}, {Divakarla}, {Dmitriev},
  {Doctor}, {Donovan}, {Dooley}, {Doravari}, {Dorrington}, {Downes}, {Drago},
  {Driggers}, {Du}, {Ducoin}, {Dudi}, {Dupej}, {Durante}, {Dwyer}, {Easter},
  {Eddolls}, {Edo}, {Effler}, {Ehrens}, {Eichholz}, {Eikenberry}, {Eisenmann},
  {Eisenstein}, {Errico}, {Essick}, {Estelles}, {Estevez}, {Etienne}, {Etzel},
  {Evans}, {Evans}, {Fafone}, {Fairhurst}, {Fan}, {Farinon}, {Farr}, {Farr},
  {Fauchon-Jones}, {Favata}, {Fays}, {Fazio}, {Fee}, {Feicht}, {Fejer}, {Feng},
  {Fernandez-Galiana}, {Ferrante}, {Ferreira}, {Ferreira}, {Fidecaro}, {Fiori},
  {Fiorucci}, {Fishbach}, {Fisher}, {Fishner}, {Fittipaldi}, {Fitz-Axen},
  {Fiumara}, {Flaminio}, {Fletcher}, {Floden}, {Flynn}, {Fong}, {Font},
  {Forsyth}, {Fournier}, {Vivanco}, {Frasca}, {Frasconi}, {Frei}, {Freise},
  {Frey}, {Frey}, {Fritschel}, {Frolov}, {Fronz{\`e}}, {Fulda}, {Fyffe},
  {Gabbard}, {Gadre}, {Gaebel}, {Gair}, {Gamba}, {Gammaitoni}, {Gaonkar},
  {Garc{\'\i}a-Quir{\'o}s}, {Garufi}, {Gateley}, {Gaudio}, {Gaur}, {Gayathri},
  {Gemme}, {Genin}, {Gennai}, {George}, {George}, {George}, {Gergely},
  {Ghonge}, {Ghosh}, {Ghosh}, {Ghosh}, {Giacomazzo}, {Giaime}, {Giardina},
  {Gibson}, {Gill}, {Glover}, {Gniesmer}, {Godwin}, {Goetz}, {Goetz},
  {Goncharov}, {Gonz{\'a}lez}, {Castro}, {Gopakumar}, {Gossan}, {Gosselin},
  {Gouaty}, {Grace}, {Grado}, {Granata}, {Grant}, {Gras}, {Grassia}, {Gray},
  {Gray}, {Greco}, {Green}, {Green}, {Gretarsson}, {Grimaldi}, {Grimm},
  {Groot}, {Grote}, {Grunewald}, {Gruning}, {Guidi}, {Gulati}, {Guo}, {Gupta},
  {Gupta}, {Gupta}, {Gustafson}, {Gustafson}, {Haegel}, {Halim}, {Hall},
  {Hall}, {Hamilton}, {Hammond}, {Haney}, {Hanke}, {Hanks}, {Hanna}, {Hannam},
  {Hannuksela}, {Hansen}, {Hanson}, {Harder}, {Hardwick}, {Haris}, {Harms},
  {Harry}, {Harry}, {Hasskew}, {Haster}, {Haughian}, {Hayes}, {Healy},
  {Heidmann}, {Heintze}, {Heitmann}, {Hellman}, {Hello}, {Hemming}, {Hendry},
  {Heng}, {Hennig}, {Heurs}, {Hild}, {Hinderer}, {Ho}, {Hochheim}, {Hofman},
  {Holgado}, {Holland}, {Holt}, {Holz}, {Hopkins}, {Horst}, {Hough}, {Howell},
  {Hoy}, {Huang}, {H{\"u}bner}, {Huerta}, {Huet}, {Hughey}, {Hui}, {Husa},
  {Huttner}, {Huynh-Dinh}, {Idzkowski}, {Iess}, {Inchauspe}, {Ingram}, {Inta},
  {Intini}, {Irwin}, {Isa}, {Isac}, {Isi}, {Iyer}, {Jacqmin}, {Jadhav}, {Jani},
  {Janthalur}, {Jaranowski}, {Jariwala}, {Jenkins}, {Jiang}, {Johnson},
  {Johnson-McDaniel}, {Jones}, {Jones}, {Jones}, {Jones}, {Jonker}, {Ju},
  {Junker}, {Kalaghatgi}, {Kalogera}, {Kamai}, {Kandhasamy}, {Kang}, {Kanner},
  {Kapadia}, {Karki}, {Kashyap}, {Kasprzack}, {Kastaun}, {Katsanevas},
  {Katsavounidis}, {Katzman}, {Kaufer}, {Kawabe}, {Keerthana},
  {K{\'e}f{\'e}lian}, {Keitel}, {Kennedy}, {Key}, {Khalili}, {Khan}, {Khan},
  {Khazanov}, {Khetan}, {Khursheed}, {Kijbunchoo}, {Kim}, {Kim}, {Kim}, {Kim},
  {Kim}, {Kim}, {Kimball}, {King}, {Kinley-Hanlon}, {Kirchhoff}, {Kissel},
  {Kleybolte}, {Klika}, {Klimenko}, {Knowles}, {Koch}, {Koehlenbeck},
  {Koekoek}, {Koley}, {Kondrashov}, {Kontos}, {Koper}, {Korobko}, {Korth},
  {Kovalam}, {Kozak}, {Kr{\"a}mer}, {Kringel}, {Krishnendu}, {Kr{\'o}lak},
  {Krupinski}, {Kuehn}, {Kumar}, {Kumar}, {Kumar}, {Kumar}, {Kuo}, {Kutynia},
  {Kwang}, {Lackey}, {Laghi}, {Lai}, {Lam}, {Landry}, {Landry}, {Lane}, {Lang},
  {Lange}, {Lantz}, {Lanza}, {Lartaux-Vollard}, {Lasky}, {Laxen}, {Lazzarini},
  {Lazzaro}, {Leaci}, {Leavey}, {Lecoeuche}, {Lee}, {Lee}, {Lee}, {Lee}, {Lee},
  {Lee}, {Lehmann}, {Lenon}, {Leroy}, {Letendre}, {Levin}, {Li}, {Li}, {Li},
  {Li}, {Li}, {Lin}, {Linde}, {Linker}, {Littenberg}, {Liu}, {Liu},
  {Llorens-Monteagudo}, {Lo}, {London}, {Longo}, {Lorenzini}, {Loriette},
  {Lormand}, {Losurdo}, {Lough}, {Lousto}, {Lovelace}, {Lower}, {Lucaccioni},
  {L{\"u}ck}, {Lumaca}, {Lundgren}, {Lynch}, {Ma}, {Macas}, {Macfoy},
  {MacInnis}, {Macleod}, {Macquet}, {Maga{\~n}a Hernandez},
  {Maga{\~n}a-Sandoval}, {Magee}, {Majorana}, {Maksimovic}, {Malik}, {Man},
  {Mandic}, {Mangano}, {Mansell}, {Manske}, {Mantovani}, {Mapelli},
  {Marchesoni}, {Marion}, {M{\'a}rka}, {M{\'a}rka}, {Markakis}, {Markosyan},
  {Markowitz}, {Maros}, {Marquina}, {Marsat}, {Martelli}, {Martin}, {Martin},
  {Martinez}, {Martynov}, {Masalehdan}, {Mason}, {Massera}, {Masserot},
  {Massinger}, {Masso-Reid}, {Mastrogiovanni}, {Matas}, {Matichard}, {Matone},
  {Mavalvala}, {McCann}, {McCarthy}, {McClelland}, {McCormick}, {McCuller},
  {McGuire}, {McIsaac}, {McIver}, {McManus}, {McRae}, {McWilliams}, {Meacher},
  {Meadors}, {Mehmet}, {Mehta}, {Meidam}, {Mejuto Villa}, {Melatos}, {Mendell},
  {Mercer}, {Mereni}, {Merfeld}, {Merilh}, {Merzougui}, {Meshkov}, {Messenger},
  {Messick}, {Messina}, {Metzdorff}, {Meyers}, {Meylahn}, {Miani}, {Miao},
  {Michel}, {Middleton}, {Milano}, {Miller}, {Millhouse}, {Mills},
  {Milovich-Goff}, {Minazzoli}, {Minenkov}, {Mishkin}, {Mishra}, {Mistry},
  {Mitra}, {Mitrofanov}, {Mitselmakher}, {Mittleman}, {Mo}, {Moffa}, {Mogushi},
  {Mohapatra}, {Molina-Ruiz}, {Mondin}, {Montani}, {Moore}, {Moraru},
  {Morawski}, {Moreno}, {Morisaki}, {Mours}, {Mow-Lowry}, {Muciaccia},
  {Mukherjee}, {Mukherjee}, {Mukherjee}, {Mukherjee}, {Mukund}, {Mullavey},
  {Munch}, {Mu{\~n}iz}, {Muratore}, {Murray}, {Nagar}, {Nardecchia},
  {Naticchioni}, {Nayak}, {Neil}, {Neilson}, {Nelemans}, {Nelson}, {Nery},
  {Neunzert}, {Nevin}, {Ng}, {Ng}, {Nguyen}, {Nguyen}, {Nichols}, {Nichols},
  {Nissanke}, {Nocera}, {North}, {Nuttall}, {Obergaulinger}, {Oberling},
  {O'Brien}, {Oganesyan}, {Ogin}, {Oh}, {Oh}, {Ohme}, {Ohta}, {Okada},
  {Oliver}, {Oppermann}, {Oram}, {O'Reilly}, {Ormiston}, {Ortega},
  {O'Shaughnessy}, {Ossokine}, {Ottaway}, {Overmier}, {Owen}, {Pace}, {Pagano},
  {Page}, {Pagliaroli}, {Pai}, {Pai}, {Palamos}, {Palashov}, {Palomba}, {Pan},
  {Panda}, {Pang}, {Pankow}, {Pannarale}, {Pant}, {Paoletti}, {Paoli},
  {Parida}, {Parker}, {Pascucci}, {Pasqualetti}, {Passaquieti}, {Passuello},
  {Patil}, {Patricelli}, {Payne}, {Pearlstone}, {Pechsiri}, {Pedersen},
  {Pedraza}, {Pedurand}, {Pele}, {Penn}, {Perego}, {Perez}, {P{\'e}rigois},
  {Perreca}, {Petermann}, {Pfeiffer}, {Phelps}, {Phukon}, {Piccinni}, {Pichot},
  {Piergiovanni}, {Pierro}, {Pillant}, {Pinard}, {Pinto}, {Pirello}, {Pitkin},
  {Plastino}, {Poggiani}, {Pong}, {Ponrathnam}, {Popolizio}, {Porter},
  {Powell}, {Prajapati}, {Prasad}, {Prasai}, {Prasanna}, {Pratten},
  {Prestegard}, {Principe}, {Prodi}, {Prokhorov}, {Punturo}, {Puppo},
  {P{\"u}rrer}, {Qi}, {Quetschke}, {Quinonez}, {Raab}, {Raaijmakers},
  {Radkins}, {Radulesco}, {Raffai}, {Raja}, {Rajan}, {Rajbhandari},
  {Rakhmanov}, {Ramirez}, {Ramos-Buades}, {Rana}, {Rao}, {Rapagnani},
  {Raymond}, {Razzano}, {Read}, {Regimbau}, {Rei}, {Reid}, {Reitze},
  {Rettegno}, {Ricci}, {Richardson}, {Richardson}, {Ricker}, {Riemenschneider},
  {Riles}, {Rizzo}, {Robertson}, {Robinet}, {Rocchi}, {Rolland}, {Rollins},
  {Roma}, {Romanelli}, {Romano}, {Romel}, {Romie}, {Rose}, {Rose}, {Rose},
  {Rosell}, {Rosi{\'n}ska}, {Rosofsky}, {Ross}, {Rowan}, {Roy}, {R{\"u}diger},
  {Ruggi}, {Rutins}, {Ryan}, {Sachdev}, {Sadecki}, {Sakellariadou}, {Salafia},
  {Salconi}, {Saleem}, {Samajdar}, {Sammut}, {Sanchez}, {Sanchez},
  {Sanchis-Gual}, {Sanders}, {Santiago}, {Santos}, {Sarin}, {Sassolas},
  {Sathyaprakash}, {Sauter}, {Savage}, {Schale}, {Scheel}, {Scheuer},
  {Schmidt}, {Schnabel}, {Schofield}, {Sch{\"o}nbeck}, {Schreiber}, {Schulte},
  {Schutz}, {Scott}, {Scott}, {Seidel}, {Sellers}, {Sengupta}, {Sennett},
  {Sentenac}, {Sequino}, {Sergeev}, {Setyawati}, {Shaddock}, {Shaffer},
  {Shahriar}, {Shaner}, {Sharma}, {Sharma}, {Shawhan}, {Shen}, {Shink},
  {Shoemaker}, {Shoemaker}, {Shukla}, {ShyamSundar}, {Siellez}, {Sieniawska},
  {Sigg}, {Singer}, {Singh}, {Singh}, {Singhal}, {Sintes}, {Sitmukhambetov},
  {Skliris}, {Slagmolen}, {Slaven-Blair}, {Smith}, {Smith}, {Somala}, {Son},
  {Soni}, {Sorazu}, {Sorrentino}, {Souradeep}, {Sowell}, {Spencer}, {Spera},
  {Srivastava}, {Srivastava}, {Staats}, {Stachie}, {Standke}, {Steer},
  {Steinke}, {Steinlechner}, {Steinlechner}, {Steinmeyer}, {Stevenson},
  {Stocks}, {Stone}, {Stops}, {Strain}, {Stratta}, {Strigin}, {Strunk},
  {Sturani}, {Stuver}, {Sudhir}, {Summerscales}, {Sun}, {Sunil}, {Sur},
  {Suresh}, {Sutton}, {Swinkels}, {Szczepa{\'n}czyk}, {Tacca}, {Tait},
  {Talbot}, {Tanner}, {Tao}, {T{\'a}pai}, {Tapia}, {Tasson}, {Taylor},
  {Tenorio}, {Terkowski}, {Thomas}, {Thomas}, {Thondapu}, {Thorne}, {Thrane},
  {Tiwari}, {Tiwari}, {Tiwari}, {Toland}, {Tonelli}, {Tornasi},
  {Torres-Forn{\'e}}, {Torrie}, {T{\"o}yr{\"a}}, {Travasso}, {Traylor},
  {Tringali}, {Tripathee}, {Trovato}, {Trozzo}, {Tsang}, {Tse}, {Tso},
  {Tsukada}, {Tsuna}, {Tsutsui}, {Tuyenbayev}, {Ueno}, {Ugolini},
  {Unnikrishnan}, {Urban}, {Usman}, {Vahlbruch}, {Vajente}, {Valdes},
  {Valentini}, {van Bakel}, {van Beuzekom}, {van den Brand}, {Van Den Broeck},
  {Vander-Hyde}, {van der Schaaf}, {VanHeijningen}, {van Veggel}, {Vardaro},
  {Varma}, {Vass}, {Vas{\'u}th}, {Vecchio}, {Vedovato}, {Veitch}, {Veitch},
  {Venkateswara}, {Venugopalan}, {Verkindt}, {Vetrano}, {Vicer{\'e}}, {Viets},
  {Vinciguerra}, {Vine}, {Vinet}, {Vitale}, {Vo}, {Vocca}, {Vorvick},
  {Vyatchanin}, {Wade}, {Wade}, {Wade}, {Walet}, {Walker}, {Wallace}, {Walsh},
  {Wang}, {Wang}, {Wang}, {Wang}, {Ward}, {Warden}, {Warner}, {Was}, {Watchi},
  {Weaver}, {Wei}, {Weinert}, {Weinstein}, {Weiss}, {Wellmann}, {Wen},
  {Wessel}, {We{\ss}els}, {Westhouse}, {Wette}, {Whelan}, {White}, {Whiting},
  {Whittle}, {Wilken}, {Williams}, {Williamson}, {Willis}, {Willke}, {Winkler},
  {Wipf}, {Wittel}, {Woan}, {Woehler}, {Wofford}, {Wright}, {Wu}, {Wysocki},
  {Xiao}, {Xu}, {Yamamoto}, {Yancey}, {Yang}, {Yang}, {Yang}, {Yap}, {Yazback},
  {Yeeles}, {Yu}, {Yu}, {Yuen}, {Zadro{\.z}ny}, {Zadro{\.z}ny}, {Zanolin},
  {Zelenova}, {Zendri}, {Zevin}, {Zhang}, {Zhang}, {Zhang}, {Zhao}, {Zhao},
  {Zhou}, {Zhou}, {Zhu}, {Zimmerman}, {Zucker}, \& {Zweizig}}]{Abbott2020low}
---. 2020{\natexlab{a}}, \apjl, 892, L3, \dodoi{10.3847/2041-8213/ab75f5}

\bibitem[{{Abbott} {et~al.}(2020{\natexlab{b}}){Abbott}, {Abbott}, {Abraham},
  {Acernese}, {Ackley}, {Adams}, {Adhikari}, {Adya}, {Affeldt}, {Agathos},
  {Agatsuma}, {Aggarwal}, {Aguiar}, {Aich}, {Aiello}, {Ain}, {Ajith}, {Akcay},
  {Allen}, {Allocca}, {Altin}, {Amato}, {Anand}, {Ananyeva}, {Anderson},
  {Anderson}, {Angelova}, {Ansoldi}, {Antier}, {Appert}, {Arai}, {Araya},
  {Areeda}, {Ar{\`e}ne}, {Arnaud}, {Aronson}, {Arun}, {Asali}, {Ascenzi},
  {Ashton}, {Aston}, {Astone}, {Aubin}, {Aufmuth}, {AultONeal}, {Austin},
  {Avendano}, {Babak}, {Bacon}, {Badaracco}, {Bader}, {Bae}, {Baer}, {Baird},
  {Baldaccini}, {Ballardin}, {Ballmer}, {Bals}, {Balsamo}, {Baltus},
  {Banagiri}, {Bankar}, {Bankar}, {Barayoga}, {Barbieri}, {Barish}, {Barker},
  {Barkett}, {Barneo}, {Barone}, {Barr}, {Barsotti}, {Barsuglia}, {Barta},
  {Bartlett}, {Bartos}, {Bassiri}, {Basti}, {Bawaj}, {Bayley}, {Bazzan},
  {B{\'e}csy}, {Bejger}, {Belahcene}, {Bell}, {Beniwal}, {Benjamin}, {Bentley},
  {Bergamin}, {Berger}, {Bergmann}, {Bernuzzi}, {Berry}, {Bersanetti},
  {Bertolini}, {Betzwieser}, {Bhandare}, {Bhandari}, {Bidler}, {Biggs},
  {Bilenko}, {Billingsley}, {Birney}, {Birnholtz}, {Biscans}, {Bischi},
  {Biscoveanu}, {Bisht}, {Bissenbayeva}, {Bitossi}, {Bizouard}, {Blackburn},
  {Blackman}, {Blair}, {Blair}, {Blair}, {Bobba}, {Bode}, {Boer}, {Boetzel},
  {Bogaert}, {Bondu}, {Bonilla}, {Bonnand}, {Booker}, {Boom}, {Bork}, {Boschi},
  {Bose}, {Bossilkov}, {Bosveld}, {Bouffanais}, {Bozzi}, {Bradaschia}, {Brady},
  {Bramley}, {Branchesi}, {Brau}, {Breschi}, {Briant}, {Briggs}, {Brighenti},
  {Brillet}, {Brinkmann}, {Brockill}, {Brooks}, {Brooks}, {Brown}, {Brunett},
  {Bruno}, {Bruntz}, {Buikema}, {Bulik}, {Bulten}, {Buonanno}, {Buscicchio},
  {Buskulic}, {Byer}, {Cabero}, {Cadonati}, {Cagnoli}, {Cahillane},
  {Calder{\'o}n Bustillo}, {Callaghan}, {Callister}, {Calloni}, {Camp},
  {Canepa}, {Cannon}, {Cao}, {Cao}, {Carapella}, {Carbognani}, {Caride},
  {Carney}, {Carullo}, {Casanueva Diaz}, {Casentini}, {Casta{\~n}eda},
  {Caudill}, {Cavagli{\`a}}, {Cavalier}, {Cavalieri}, {Cella},
  {Cerd{\'a}-Dur{\'a}n}, {Cesarini}, {Chaibi}, {Chakravarti}, {Chan}, {Chan},
  {Chandra}, {Chao}, {Charlton}, {Chase}, {Chassande-Mottin}, {Chatterjee},
  {Chaturvedi}, {Chatziioannou}, {Chen}, {Chen}, {Chen}, {Cheng}, {Cheong},
  {Chia}, {Chiadini}, {Chierici}, {Chincarini}, {Chiummo}, {Cho}, {Cho}, {Cho},
  {Christensen}, {Chu}, {Chua}, {Chung}, {Chung}, {Ciani}, {Ciecielag},
  {Cie{\'s}lar}, {Ciobanu}, {Ciolfi}, {Cipriano}, {Cirone}, {Clara}, {Clark},
  {Clearwater}, {Clesse}, {Cleva}, {Coccia}, {Cohadon}, {Cohen}, {Colleoni},
  {Collette}, {Collins}, {Colpi}, {Constancio}, {Conti}, {Cooper}, {Corban},
  {Corbitt}, {Cordero-Carri{\'o}n}, {Corezzi}, {Corley}, {Cornish}, {Corre},
  {Corsi}, {Cortese}, {Costa}, {Cotesta}, {Coughlin}, {Coughlin}, {Coulon},
  {Countryman}, {Couvares}, {Covas}, {Coward}, {Cowart}, {Coyne}, {Coyne},
  {Creighton}, {Creighton}, {Cripe}, {Croquette}, {Crowder}, {Cudell},
  {Cullen}, {Cumming}, {Cummings}, {Cunningham}, {Cuoco}, {Curylo}, {Canton},
  {D{\'a}lya}, {Dana}, {Daneshgaran-Bajastani}, {D'Angelo}, {Danilishin},
  {D'Antonio}, {Danzmann}, {Darsow-Fromm}, {Dasgupta}, {Datrier}, {Dattilo},
  {Dave}, {Davier}, {Davies}, {Davis}, {Daw}, {DeBra}, {Deenadayalan},
  {Degallaix}, {De Laurentis}, {Del{\'e}glise}, {Delfavero}, {De Lillo}, {Del
  Pozzo}, {DeMarchi}, {D'Emilio}, {Demos}, {Dent}, {De Pietri}, {De Rosa}, {De
  Rossi}, {DeSalvo}, {de Varona}, {Dhurandhar}, {D{\'\i}az}, {Diaz-Ortiz},
  {Dietrich}, {Di Fiore}, {Di Fronzo}, {Di Giorgio}, {Di Giovanni}, {Di
  Giovanni}, {Di Girolamo}, {Di Lieto}, {Ding}, {Di Pace}, {Di Palma}, {Di
  Renzo}, {Divakarla}, {Dmitriev}, {Doctor}, {Donovan}, {Dooley}, {Doravari},
  {Dorrington}, {Downes}, {Drago}, {Driggers}, {Du}, {Ducoin}, {Dupej},
  {Durante}, {D'Urso}, {Dwyer}, {Easter}, {Eddolls}, {Edelman}, {Edo}, {Edy},
  {Effler}, {Ehrens}, {Eichholz}, {Eikenberry}, {Eisenmann}, {Eisenstein},
  {Ejlli}, {Errico}, {Essick}, {Estelles}, {Estevez}, {Etienne}, {Etzel},
  {Evans}, {Evans}, {Ewing}, {Fafone}, {Fairhurst}, {Fan}, {Farinon}, {Farr},
  {Farr}, {Fauchon-Jones}, {Favata}, {Fays}, {Fazio}, {Feicht}, {Fejer},
  {Feng}, {Fenyvesi}, {Ferguson}, {Fernandez-Galiana}, {Ferrante}, {Ferreira},
  {Ferreira}, {Fidecaro}, {Fiori}, {Fiorucci}, {Fishbach}, {Fisher},
  {Fittipaldi}, {Fitz-Axen}, {Fiumara}, {Flaminio}, {Floden}, {Flynn}, {Fong},
  {Font}, {Forsyth}, {Fournier}, {Frasca}, {Frasconi}, {Frei}, {Freise},
  {Frey}, {Frey}, {Fritschel}, {Frolov}, {Fronz{\`e}}, {Fulda}, {Fyffe},
  {Gabbard}, {Gadre}, {Gaebel}, {Gair}, {Galaudage}, {Ganapathy}, {Ganguly},
  {Gaonkar}, {Garc{\'\i}a-Quir{\'o}s}, {Garufi}, {Gateley}, {Gaudio},
  {Gayathri}, {Gemme}, {Genin}, {Gennai}, {George}, {George}, {Gergely},
  {Ghonge}, {Ghosh}, {Ghosh}, {Ghosh}, {Giacomazzo}, {Giaime}, {Giardina},
  {Gibson}, {Gier}, {Gill}, {Glanzer}, {Gniesmer}, {Godwin}, {Goetz}, {Goetz},
  {Gohlke}, {Goncharov}, {Gonz{\'a}lez}, {Gopakumar}, {Gossan}, {Gosselin},
  {Gouaty}, {Grace}, {Grado}, {Granata}, {Grant}, {Gras}, {Grassia}, {Gray},
  {Gray}, {Greco}, {Green}, {Green}, {Gretarsson}, {Griggs}, {Grignani},
  {Grimaldi}, {Grimm}, {Grote}, {Grunewald}, {Gruning}, {Guidi}, {Guimaraes},
  {Guix{\'e}}, {Gulati}, {Guo}, {Gupta}, {Gupta}, {Gupta}, {Gustafson},
  {Gustafson}, {Haegel}, {Halim}, {Hall}, {Hamilton}, {Hammond}, {Haney},
  {Hanke}, {Hanks}, {Hanna}, {Hannam}, {Hannuksela}, {Hansen}, {Hanson},
  {Harder}, {Hardwick}, {Haris}, {Harms}, {Harry}, {Harry}, {Hasskew},
  {Haster}, {Haughian}, {Hayes}, {Healy}, {Heidmann}, {Heintze}, {Heinze},
  {Heitmann}, {Hellman}, {Hello}, {Hemming}, {Hendry}, {Heng}, {Hennes},
  {Hennig}, {Heurs}, {Hild}, {Hinderer}, {Hoback}, {Hochheim}, {Hofgard},
  {Hofman}, {Holgado}, {Holland}, {Holt}, {Holz}, {Hopkins}, {Horst}, {Hough},
  {Howell}, {Hoy}, {Huang}, {H{\"u}bner}, {Huerta}, {Huet}, {Hughey}, {Hui},
  {Husa}, {Huttner}, {Huxford}, {Huynh-Dinh}, {Idzkowski}, {Iess}, {Inchauspe},
  {Ingram}, {Intini}, {Isac}, {Isi}, {Iyer}, {Jacqmin}, {Jadhav}, {Jadhav},
  {James}, {Jani}, {Janthalur}, {Jaranowski}, {Jariwala}, {Jaume}, {Jenkins},
  {Jiang}, {Johns}, {Johnson-McDaniel}, {Jones}, {Jones}, {Jones}, {Jones},
  {Jones}, {Jonker}, {Ju}, {Junker}, {Kalaghatgi}, {Kalogera}, {Kamai},
  {Kandhasamy}, {Kang}, {Kanner}, {Kapadia}, {Karki}, {Kashyap}, {Kasprzack},
  {Kastaun}, {Katsanevas}, {Katsavounidis}, {Katzman}, {Kaufer}, {Kawabe},
  {K{\'e}f{\'e}lian}, {Keitel}, {Keivani}, {Kennedy}, {Key}, {Khadka},
  {Khalili}, {Khan}, {Khan}, {Khan}, {Khazanov}, {Khetan}, {Khursheed},
  {Kijbunchoo}, {Kim}, {Kim}, {Kim}, {Kim}, {Kim}, {Kim}, {Kim}, {Kimball},
  {King}, {Kinley-Hanlon}, {Kirchhoff}, {Kissel}, {Kleybolte}, {Klimenko},
  {Knowles}, {Knyazev}, {Koch}, {Koehlenbeck}, {Koekoek}, {Koley},
  {Kondrashov}, {Kontos}, {Koper}, {Korobko}, {Korth}, {Kovalam}, {Kozak},
  {Kringel}, {Krishnendu}, {Kr{\'o}lak}, {Krupinski}, {Kuehn}, {Kumar},
  {Kumar}, {Kumar}, {Kumar}, {Kumar}, {Kuo}, {Kutynia}, {Lackey}, {Laghi},
  {Lalande}, {Lam}, {Lamberts}, {Landry}, {Lane}, {Lang}, {Lange}, {Lantz},
  {Lanza}, {La Rosa}, {Lartaux-Vollard}, {Lasky}, {Laxen}, {Lazzarini},
  {Lazzaro}, {Leaci}, {Leavey}, {Lecoeuche}, {Lee}, {Lee}, {Lee}, {Lee}, {Lee},
  {Lehmann}, {Leroy}, {Letendre}, {Levin}, {Li}, {Li}, {li}, {Li}, {Li},
  {Linde}, {Linker}, {Linley}, {Littenberg}, {Liu}, {Liu},
  {Llorens-Monteagudo}, {Lo}, {Lockwood}, {London}, {Longo}, {Lorenzini},
  {Loriette}, {Lormand}, {Losurdo}, {Lough}, {Lousto}, {Lovelace}, {L{\"u}ck},
  {Lumaca}, {Lundgren}, {Ma}, {Macas}, {Macfoy}, {MacInnis}, {Macleod},
  {MacMillan}, {Macquet}, {Maga{\~n}a Hernandez}, {Maga{\~n}a-Sandoval},
  {Magee}, {Majorana}, {Maksimovic}, {Malik}, {Man}, {Mandic}, {Mangano},
  {Mansell}, {Manske}, {Mantovani}, {Mapelli}, {Marchesoni}, {Marion},
  {M{\'a}rka}, {M{\'a}rka}, {Markakis}, {Markosyan}, {Markowitz}, {Maros},
  {Marquina}, {Marsat}, {Martelli}, {Martin}, {Martin}, {Martinez}, {Martynov},
  {Masalehdan}, {Mason}, {Massera}, {Masserot}, {Massinger}, {Masso-Reid},
  {Mastrogiovanni}, {Matas}, {Matichard}, {Mavalvala}, {Maynard}, {McCann},
  {McCarthy}, {McClelland}, {McCormick}, {McCuller}, {McGuire}, {McIsaac},
  {McIver}, {McManus}, {McRae}, {McWilliams}, {Meacher}, {Meadors}, {Mehmet},
  {Mehta}, {Mejuto Villa}, {Melatos}, {Mendell}, {Mercer}, {Mereni}, {Merfeld},
  {Merilh}, {Merritt}, {Merzougui}, {Meshkov}, {Messenger}, {Messick},
  {Metzdorff}, {Meyers}, {Meylahn}, {Mhaske}, {Miani}, {Miao}, {Michaloliakos},
  {Michel}, {Middleton}, {Milano}, {Miller}, {Millhouse}, {Mills}, {Milotti},
  {Milovich-Goff}, {Minazzoli}, {Minenkov}, {Mishkin}, {Mishra}, {Mistry},
  {Mitra}, {Mitrofanov}, {Mitselmakher}, {Mittleman}, {Mo}, {Mogushi},
  {Mohapatra}, {Mohite}, {Molina-Ruiz}, {Mondin}, {Montani}, {Moore}, {Moraru},
  {Morawski}, {Moreno}, {Morisaki}, {Mours}, {Mow-Lowry}, {Mozzon},
  {Muciaccia}, {Mukherjee}, {Mukherjee}, {Mukherjee}, {Mukherjee}, {Mukund},
  {Mullavey}, {Munch}, {Mu{\~n}iz}, {Murray}, {Nagar}, {Nardecchia},
  {Naticchioni}, {Nayak}, {Neil}, {Neilson}, {Nelemans}, {Nelson}, {Nery},
  {Neunzert}, {Ng}, {Ng}, {Nguyen}, {Nguyen}, {Nichols}, {Nichols}, {Nissanke},
  {Nitz}, {Nocera}, {Noh}, {North}, {Nothard}, {Nuttall}, {Oberling},
  {O'Brien}, {Oganesyan}, {Ogin}, {Oh}, {Oh}, {Ohme}, {Ohta}, {Okada},
  {Oliver}, {Olivetto}, {Oppermann}, {Oram}, {O'Reilly}, {Ormiston}, {Ortega},
  {O'Shaughnessy}, {Ossokine}, {Osthelder}, {Ottaway}, {Overmier}, {Owen},
  {Pace}, {Pagano}, {Page}, {Pagliaroli}, {Pai}, {Pai}, {Palamos}, {Palashov},
  {Palomba}, {Pan}, {Panda}, {Pang}, {Pankow}, {Pannarale}, {Pant}, {Paoletti},
  {Paoli}, {Parida}, {Parker}, {Pascucci}, {Pasqualetti}, {Passaquieti},
  {Passuello}, {Patricelli}, {Payne}, {Pearlstone}, {Pechsiri}, {Pedersen},
  {Pedraza}, {Pele}, {Penn}, {Perego}, {Perez}, {P{\'e}rigois}, {Perreca},
  {Perri{\`e}s}, {Petermann}, {Pfeiffer}, {Phelps}, {Phukon}, {Piccinni},
  {Pichot}, {Piendibene}, {Piergiovanni}, {Pierro}, {Pillant}, {Pinard},
  {Pinto}, {Piotrzkowski}, {Pirello}, {Pitkin}, {Plastino}, {Poggiani}, {Pong},
  {Ponrathnam}, {Popolizio}, {Porter}, {Powell}, {Prajapati}, {Prasai},
  {Prasanna}, {Pratten}, {Prestegard}, {Principe}, {Prodi}, {Prokhorov},
  {Punturo}, {Puppo}, {P{\"u}rrer}, {Qi}, {Quetschke}, {Quinonez}, {Raab},
  {Raaijmakers}, {Radkins}, {Radulesco}, {Raffai}, {Rafferty}, {Raja}, {Rajan},
  {Rajbhandari}, {Rakhmanov}, {Ramirez}, {Ramos-Buades}, {Rana}, {Rao},
  {Rapagnani}, {Raymond}, {Razzano}, {Read}, {Regimbau}, {Rei}, {Reid},
  {Reitze}, {Rettegno}, {Ricci}, {Richardson}, {Richardson}, {Ricker},
  {Riemenschneider}, {Riles}, {Rizzo}, {Robertson}, {Robinet}, {Rocchi},
  {Rodriguez-Soto}, {Rolland}, {Rollins}, {Roma}, {Romanelli}, {Romano},
  {Romel}, {Romero-Shaw}, {Romie}, {Rose}, {Rose}, {Rose}, {Rosi{\'n}ska},
  {Rosofsky}, {Ross}, {Rowan}, {Rowlinson}, {Roy}, {Roy}, {Roy}, {Ruggi},
  {Rutins}, {Ryan}, {Sachdev}, {Sadecki}, {Sakellariadou}, {Salafia},
  {Salconi}, {Saleem}, {Salemi}, {Samajdar}, {Sanchez}, {Sanchez},
  {Sanchis-Gual}, {Sanders}, {Santiago}, {Santos}, {Sarin}, {Sassolas},
  {Sathyaprakash}, {Sauter}, {Savage}, {Savant}, {Sawant}, {Sayah}, {Schaetzl},
  {Schale}, {Scheel}, {Scheuer}, {Schmidt}, {Schnabel}, {Schofield},
  {Sch{\"o}nbeck}, {Schreiber}, {Schulte}, {Schutz}, {Schwarm}, {Schwartz},
  {Scott}, {Scott}, {Seidel}, {Sellers}, {Sengupta}, {Sennett}, {Sentenac},
  {Sequino}, {Sergeev}, {Setyawati}, {Shaddock}, {Shaffer}, {Sharifi},
  {Shahriar}, {Sharma}, {Sharma}, {Shawhan}, {Shen}, {Shikauchi}, {Shink},
  {Shoemaker}, {Shoemaker}, {Shukla}, {ShyamSundar}, {Siellez}, {Sieniawska},
  {Sigg}, {Singer}, {Singh}, {Singh}, {Singha}, {Singhal}, {Sintes}, {Sipala},
  {Skliris}, {Slagmolen}, {Slaven-Blair}, {Smetana}, {Smith}, {Smith},
  {Somala}, {Son}, {Soni}, {Sorazu}, {Sordini}, {Sorrentino}, {Souradeep},
  {Sowell}, {Spencer}, {Spera}, {Srivastava}, {Srivastava}, {Staats},
  {Stachie}, {Standke}, {Steer}, {Steinke}, {Steinlechner}, {Steinlechner},
  {Steinmeyer}, {Stevenson}, {Stocks}, {Stops}, {Stover}, {Strain}, {Stratta},
  {Strunk}, {Sturani}, {Stuver}, {Sudhagar}, {Sudhir}, {Summerscales}, {Sun},
  {Sunil}, {Sur}, {Suresh}, {Sutton}, {Swinkels}, {Szczepa{\'n}czyk}, {Tacca},
  {Tait}, {Talbot}, {Tanasijczuk}, {Tanner}, {Tao}, {T{\'a}pai}, {Tapia},
  {Tapia San Martin}, {Tasson}, {Taylor}, {Tenorio}, {Terkowski},
  {Thirugnanasambandam}, {Thomas}, {Thomas}, {Thompson}, {Thondapu}, {Thorne},
  {Thrane}, {Tinsman}, {Saravanan}, {Tiwari}, {Tiwari}, {Tiwari}, {Toland},
  {Tonelli}, {Tornasi}, {Torres-Forn{\'e}}, {Torrie}, {Tosta e Melo},
  {T{\"o}yr{\"a}}, {Travasso}, {Traylor}, {Tringali}, {Tripathee}, {Trovato},
  {Trudeau}, {Tsang}, {Tse}, {Tso}, {Tsukada}, {Tsuna}, {Tsutsui}, {Turconi},
  {Ubhi}, {Udall}, {Ueno}, {Ugolini}, {Unnikrishnan}, {Urban}, {Usman},
  {Utina}, {Vahlbruch}, {Vajente}, {Valdes}, {Valentini}, {van Bakel}, {van
  Beuzekom}, {van den Brand}, {Van Den Broeck}, {Vander-Hyde}, {van der
  Schaaf}, {Van Heijningen}, {van Veggel}, {Vardaro}, {Varma}, {Vass},
  {Vas{\'u}th}, {Vecchio}, {Vedovato}, {Veitch}, {Veitch}, {Venkateswara},
  {Venugopalan}, {Verkindt}, {Veske}, {Vetrano}, {Vicer{\'e}}, {Viets},
  {Vinciguerra}, {Vine}, {Vinet}, {Vitale}, {Vivanco}, {Vo}, {Vocca},
  {Vorvick}, {Vyatchanin}, {Wade}, {Wade}, {Wade}, {Walet}, {Walker},
  {Wallace}, {Wallace}, {Walsh}, {Wang}, {Wang}, {Wang}, {Ward}, {Warden},
  {Warner}, {Was}, {Watchi}, {Weaver}, {Wei}, {Weinert}, {Weinstein}, {Weiss},
  {Wellmann}, {Wen}, {We{\ss}els}, {Westhouse}, {Wette}, {Whelan}, {Whiting},
  {Whittle}, {Wilken}, {Williams}, {Willis}, {Willke}, {Winkler}, {Wipf},
  {Wittel}, {Woan}, {Woehler}, {Wofford}, {Wong}, {Wright}, {Wu}, {Wysocki},
  {Xiao}, {Yamamoto}, {Yang}, {Yang}, {Yang}, {Yap}, {Yazback}, {Yeeles}, {Yu},
  {Yu}, {Yuen}, {Zadro{\.Z}ny}, {Zadro{\.Z}ny}, {Zanolin}, {Zelenova},
  {Zendri}, {Zevin}, {Zhang}, {Zhang}, {Zhang}, {Zhao}, {Zhao}, {Zhou}, {Zhou},
  {Zhu}, {Zimmerman}, {Zucker}, {Zweizig}, {LIGO Scientific Collaboration}, \&
  {Virgo Collaboration}}]{Abbott2020high}
{Abbott}, R., {Abbott}, T.~D., {Abraham}, S., {et~al.} 2020{\natexlab{b}},
  \prl, 125, 101102, \dodoi{10.1103/PhysRevLett.125.101102}

\bibitem[{{Aloy} {et~al.}(2000){Aloy}, {M{\"u}ller}, {Ib{\'a}{\~n}ez},
  {Mart{\'\i}}, \& {MacFadyen}}]{Aloy2000}
{Aloy}, M.~A., {M{\"u}ller}, E., {Ib{\'a}{\~n}ez}, J.~M., {Mart{\'\i}}, J.~M.,
  \& {MacFadyen}, A. 2000, \apjl, 531, L119, \dodoi{10.1086/312537}

\bibitem[{{Artymowicz} {et~al.}(1993){Artymowicz}, {Lin}, \&
  {Wampler}}]{Artymowicz1993}
{Artymowicz}, P., {Lin}, D.~N.~C., \& {Wampler}, E.~J. 1993, \apj, 409, 592,
  \dodoi{10.1086/172690}

\bibitem[{{Bellovary} {et~al.}(2016){Bellovary}, {Mac Low}, {McKernan}, \&
  {Ford}}]{Bellovary2016}
{Bellovary}, J.~M., {Mac Low}, M.-M., {McKernan}, B., \& {Ford}, K.~E.~S. 2016,
  \apjl, 819, L17, \dodoi{10.3847/2041-8205/819/2/L17}

\bibitem[{{Bromberg} \& {Levinson}(2007)}]{Bromberg2007}
{Bromberg}, O., \& {Levinson}, A. 2007, \apj, 671, 678, \dodoi{10.1086/522668}

\bibitem[{{Callister} {et~al.}(2021){Callister}, {Haster}, {Ng}, {Vitale}, \&
  {Farr}}]{Callister2021}
{Callister}, T.~A., {Haster}, C.-J., {Ng}, K. K.~Y., {Vitale}, S., \& {Farr},
  W.~M. 2021, \apjl, 922, L5, \dodoi{10.3847/2041-8213/ac2ccc}

\bibitem[{{Cantiello} {et~al.}(2021){Cantiello}, {Jermyn}, \&
  {Lin}}]{Cantiello2021}
{Cantiello}, M., {Jermyn}, A.~S., \& {Lin}, D. N.~C. 2021, \apj, 910, 94,
  \dodoi{10.3847/1538-4357/abdf4f}

\bibitem[{{Chhotray} \& {Lazzati}(2015)}]{Chhotray2015}
{Chhotray}, A., \& {Lazzati}, D. 2015, \apj, 802, 132,
  \dodoi{10.1088/0004-637X/802/2/132}

\bibitem[{{Della Valle} {et~al.}(2006){Della Valle}, {Chincarini}, {Panagia},
  {Tagliaferri}, {Malesani}, {Testa}, {Fugazza}, {Campana}, {Covino},
  {Mangano}, {Antonelli}, {D'Avanzo}, {Hurley}, {Mirabel}, {Pellizza},
  {Piranomonte}, \& {Stella}}]{DellaValla2006}
{Della Valle}, M., {Chincarini}, G., {Panagia}, N., {et~al.} 2006, \nat, 444,
  1050, \dodoi{10.1038/nature05374}

\bibitem[{{Dittmann} {et~al.}(2021){Dittmann}, {Cantiello}, \&
  {Jermyn}}]{Dittmann2021}
{Dittmann}, A.~J., {Cantiello}, M., \& {Jermyn}, A.~S. 2021, \apj, 916, 48,
  \dodoi{10.3847/1538-4357/ac042c}

\bibitem[{{Dittmann} \& {Miller}(2020)}]{Dittmann2020}
{Dittmann}, A.~J., \& {Miller}, M.~C. 2020, \mnras, 493, 3732,
  \dodoi{10.1093/mnras/staa463}

\bibitem[{{Fabj} {et~al.}(2020){Fabj}, {Nasim}, {Caban}, {Ford}, {McKernan}, \&
  {Bellovary}}]{Fabj2020}
{Fabj}, G., {Nasim}, S.~S., {Caban}, F., {et~al.} 2020, \mnras, 499, 2608,
  \dodoi{10.1093/mnras/staa3004}

\bibitem[{{Foreman-Mackey} {et~al.}(2013){Foreman-Mackey}, {Hogg}, {Lang}, \&
  {Goodman}}]{Foreman2013}
{Foreman-Mackey}, D., {Hogg}, D.~W., {Lang}, D., \& {Goodman}, J. 2013, \pasp,
  125, 306, \dodoi{10.1086/670067}

\bibitem[{{Fynbo} {et~al.}(2006){Fynbo}, {Watson}, {Th{\"o}ne}, {Sollerman},
  {Bloom}, {Davis}, {Hjorth}, {Jakobsson}, {J{\o}rgensen}, {Graham},
  {Fruchter}, {Bersier}, {Kewley}, {Cassan}, {Castro Cer{\'o}n}, {Foley},
  {Gorosabel}, {Hinse}, {Horne}, {Jensen}, {Klose}, {Kocevski}, {Marquette},
  {Perley}, {Ramirez-Ruiz}, {Stritzinger}, {Vreeswijk}, {Wijers}, {Woller},
  {Xu}, \& {Zub}}]{Fynbo2006}
{Fynbo}, J. P.~U., {Watson}, D., {Th{\"o}ne}, C.~C., {et~al.} 2006, \nat, 444,
  1047, \dodoi{10.1038/nature05375}

\bibitem[{{Gal-Yam} {et~al.}(2006){Gal-Yam}, {Fox}, {Price}, {Ofek}, {Davis},
  {Leonard}, {Soderberg}, {Schmidt}, {Lewis}, {Peterson}, {Kulkarni}, {Berger},
  {Cenko}, {Sari}, {Sharon}, {Frail}, {Moon}, {Brown}, {Cucchiara}, {Harrison},
  {Piran}, {Persson}, {McCarthy}, {Penprase}, {Chevalier}, \&
  {MacFadyen}}]{Gal-Yam2006}
{Gal-Yam}, A., {Fox}, D.~B., {Price}, P.~A., {et~al.} 2006, \nat, 444, 1053,
  \dodoi{10.1038/nature05373}

\bibitem[{{Gehrels} {et~al.}(2006){Gehrels}, {Norris}, {Barthelmy}, {Granot},
  {Kaneko}, {Kouveliotou}, {Markwardt}, {M{\'e}sz{\'a}ros}, {Nakar}, {Nousek},
  {O'Brien}, {Page}, {Palmer}, {Parsons}, {Roming}, {Sakamoto}, {Sarazin},
  {Schady}, {Stamatikos}, \& {Woosley}}]{Gehrels2006}
{Gehrels}, N., {Norris}, J.~P., {Barthelmy}, S.~D., {et~al.} 2006, \nat, 444,
  1044, \dodoi{10.1038/nature05376}

\bibitem[{{Ghisellini} {et~al.}(1998){Ghisellini}, {Haardt}, \&
  {Svensson}}]{Ghisellini1998}
{Ghisellini}, G., {Haardt}, F., \& {Svensson}, R. 1998, \mnras, 297, 348,
  \dodoi{10.1046/j.1365-8711.1998.01442.x}

\bibitem[{{Giannios}(2006)}]{Giannios2006}
{Giannios}, D. 2006, \aap, 457, 763, \dodoi{10.1051/0004-6361:20065000}

\bibitem[{{Giannios} \& {Spruit}(2007)}]{Giannios2007}
{Giannios}, D., \& {Spruit}, H.~C. 2007, \aap, 469, 1,
  \dodoi{10.1051/0004-6361:20066739}

\bibitem[{{Gompertz} {et~al.}(2023){Gompertz}, {Ravasio}, {Nicholl}, {Levan},
  {Metzger}, {Oates}, {Lamb}, {Fong}, {Malesani}, {Rastinejad}, {Tanvir},
  {Evans}, {Jonker}, {Page}, \& {Pe'er}}]{Gompertz23}
{Gompertz}, B.~P., {Ravasio}, M.~E., {Nicholl}, M., {et~al.} 2023, Nature
  Astronomy, 7, 67, \dodoi{10.1038/s41550-022-01819-4}

\bibitem[{{Goodman}(2003)}]{Goodman2003}
{Goodman}, J. 2003, \mnras, 339, 937, \dodoi{10.1046/j.1365-8711.2003.06241.x}

\bibitem[{{Gottlieb} \& {Nakar}(2022)}]{Gottlieb2022}
{Gottlieb}, O., \& {Nakar}, E. 2022, \mnras, 517, 1640,
  \dodoi{10.1093/mnras/stac2699}

\bibitem[{{Gottlieb} {et~al.}(2021){Gottlieb}, {Nakar}, \&
  {Bromberg}}]{Gottlieb2021}
{Gottlieb}, O., {Nakar}, E., \& {Bromberg}, O. 2021, \mnras, 500, 3511,
  \dodoi{10.1093/mnras/staa3501}

\bibitem[{{Hopkins} {et~al.}(2012){Hopkins}, {Hernquist}, {Hayward}, \&
  {Narayanan}}]{Hopkins2012}
{Hopkins}, P.~F., {Hernquist}, L., {Hayward}, C.~C., \& {Narayanan}, D. 2012,
  \mnras, 425, 1121, \dodoi{10.1111/j.1365-2966.2012.21449.x}

\bibitem[{{Ito} {et~al.}(2021){Ito}, {Just}, {Takei}, \& {Nagataki}}]{Ito2021}
{Ito}, H., {Just}, O., {Takei}, Y., \& {Nagataki}, S. 2021, \apj, 918, 59,
  \dodoi{10.3847/1538-4357/ac0cf9}

\bibitem[{{Ito} {et~al.}(2015){Ito}, {Matsumoto}, {Nagataki}, {Warren}, \&
  {Barkov}}]{Ito2015}
{Ito}, H., {Matsumoto}, J., {Nagataki}, S., {Warren}, D.~C., \& {Barkov}, M.~V.
  2015, \apjl, 814, L29, \dodoi{10.1088/2041-8205/814/2/L29}

\bibitem[{{Jermyn} {et~al.}(2021){Jermyn}, {Dittmann}, {Cantiello}, \&
  {Perna}}]{Jermyn2021}
{Jermyn}, A.~S., {Dittmann}, A.~J., {Cantiello}, M., \& {Perna}, R. 2021, \apj,
  914, 105, \dodoi{10.3847/1538-4357/abfb67}

\bibitem[{{Kinney} {et~al.}(2000){Kinney}, {Schmitt}, {Clarke}, {Pringle},
  {Ulvestad}, \& {Antonucci}}]{kinney2000}
{Kinney}, A.~L., {Schmitt}, H.~R., {Clarke}, C.~J., {et~al.} 2000, \apj, 537,
  152, \dodoi{10.1086/309016}

\bibitem[{{Lazzati}(2016)}]{Lazzati2016}
{Lazzati}, D. 2016, \apj, 829, 76, \dodoi{10.3847/0004-637X/829/2/76}

\bibitem[{{Lazzati} {et~al.}(2020){Lazzati}, {Ciolfi}, \&
  {Perna}}]{Lazzati2020}
{Lazzati}, D., {Ciolfi}, R., \& {Perna}, R. 2020, \apj, 898, 59,
  \dodoi{10.3847/1538-4357/ab9a44}

\bibitem[{{Lazzati} \& {Perna}(2019)}]{Lazzati2019}
{Lazzati}, D., \& {Perna}, R. 2019, \apj, 881, 89,
  \dodoi{10.3847/1538-4357/ab2e06}

\bibitem[{{Lazzati} {et~al.}(2018){Lazzati}, {Perna}, {Morsony},
  {Lopez-Camara}, {Cantiello}, {Ciolfi}, {Giacomazzo}, \&
  {Workman}}]{Lazzati2018}
{Lazzati}, D., {Perna}, R., {Morsony}, B.~J., {et~al.} 2018, \prl, 120, 241103,
  \dodoi{10.1103/PhysRevLett.120.241103}

\bibitem[{{Lazzati} {et~al.}(2022){Lazzati}, {Soares}, \&
  {Perna}}]{Lazzati2022}
{Lazzati}, D., {Soares}, G., \& {Perna}, R. 2022, \apjl, 938, L18,
  \dodoi{10.3847/2041-8213/ac98ad}

\bibitem[{{Levan} {et~al.}(2023){Levan}, {Malesani}, {Gompertz}, {three},
  {four}, {ciccio}, \& {panza}}]{Levan2023}
{Levan}, A.~J., {Malesani}, D., {Gompertz}, B., {et~al.} 2023, Accepted for
  publication in Nature Astronomy,
  \dodoi{https://doi.org/10.21203/rs.3.rs-2298504/v1}

\bibitem[{{MacFadyen} \& {Woosley}(1999)}]{MacFadyen1999}
{MacFadyen}, A.~I., \& {Woosley}, S.~E. 1999, \apj, 524, 262,
  \dodoi{10.1086/307790}

\bibitem[{{Matzner}(2003)}]{Matzner2003}
{Matzner}, C.~D. 2003, \mnras, 345, 575,
  \dodoi{10.1046/j.1365-8711.2003.06969.x}

\bibitem[{{McKernan} {et~al.}(2021){McKernan}, {Ford}, {Callister}, {Farr},
  {O'Shaughnessy}, {Smith}, {Thrane}, \& {Vajpeyi}}]{McKernan2021}
{McKernan}, B., {Ford}, K.~E.~S., {Callister}, T., {et~al.} 2021, arXiv
  e-prints, arXiv:2107.07551.
\newblock \doarXiv{2107.07551}

\bibitem[{{McKernan} {et~al.}(2020){McKernan}, {Ford}, \&
  {O'Shaughnessy}}]{McKernan2020}
{McKernan}, B., {Ford}, K.~E.~S., \& {O'Shaughnessy}, R. 2020, \mnras, 498,
  4088, \dodoi{10.1093/mnras/staa2681}

\bibitem[{{Mei} {et~al.}(2022){Mei}, {Banerjee}, {Oganesyan}, {Salafia},
  {Giarratana}, {Branchesi}, {D'Avanzo}, {Campana}, {Ghirlanda}, {Ronchini},
  {Shukla}, \& {Tiwari}}]{Mei2022}
{Mei}, A., {Banerjee}, B., {Oganesyan}, G., {et~al.} 2022, \nat, 612, 236,
  \dodoi{10.1038/s41586-022-05404-7}

\bibitem[{{Morsony} {et~al.}(2007){Morsony}, {Lazzati}, \&
  {Begelman}}]{Morsony2007}
{Morsony}, B.~J., {Lazzati}, D., \& {Begelman}, M.~C. 2007, \apj, 665, 569,
  \dodoi{10.1086/519483}

\bibitem[{{Murguia-Berthier} {et~al.}(2014){Murguia-Berthier}, {Montes},
  {Ramirez-Ruiz}, {De Colle}, \& {Lee}}]{Murguia2014}
{Murguia-Berthier}, A., {Montes}, G., {Ramirez-Ruiz}, E., {De Colle}, F., \&
  {Lee}, W.~H. 2014, \apjl, 788, L8, \dodoi{10.1088/2041-8205/788/1/L8}

\bibitem[{{Nagakura} {et~al.}(2014){Nagakura}, {Hotokezaka}, {Sekiguchi},
  {Shibata}, \& {Ioka}}]{Nagakura2014}
{Nagakura}, H., {Hotokezaka}, K., {Sekiguchi}, Y., {Shibata}, M., \& {Ioka}, K.
  2014, \apjl, 784, L28, \dodoi{10.1088/2041-8205/784/2/L28}

\bibitem[{{Parsotan} \& {Lazzati}(2018)}]{Parsotan2018}
{Parsotan}, T., \& {Lazzati}, D. 2018, \apj, 853, 8,
  \dodoi{10.3847/1538-4357/aaa087}

\bibitem[{{Pe'er} {et~al.}(2005){Pe'er}, {M{\'e}sz{\'a}ros}, \&
  {Rees}}]{Peer2005}
{Pe'er}, A., {M{\'e}sz{\'a}ros}, P., \& {Rees}, M.~J. 2005, \apj, 635, 476,
  \dodoi{10.1086/497360}

\bibitem[{{Pe'er} {et~al.}(2006){Pe'er}, {M{\'e}sz{\'a}ros}, \&
  {Rees}}]{Peer2006}
---. 2006, \apj, 642, 995, \dodoi{10.1086/501424}

\bibitem[{{Perna} {et~al.}(2021{\natexlab{a}}){Perna}, {Lazzati}, \&
  {Cantiello}}]{Perna2021}
{Perna}, R., {Lazzati}, D., \& {Cantiello}, M. 2021{\natexlab{a}}, \apjl, 906,
  L7, \dodoi{10.3847/2041-8213/abd319}

\bibitem[{{Perna} {et~al.}(2021{\natexlab{b}}){Perna}, {Tagawa}, {Haiman}, \&
  {Bartos}}]{Perna2021b}
{Perna}, R., {Tagawa}, H., {Haiman}, Z., \& {Bartos}, I. 2021{\natexlab{b}},
  \apj, 915, 10, \dodoi{10.3847/1538-4357/abfdb4}

\bibitem[{{Piran}(2004)}]{Piran2004}
{Piran}, T. 2004, Reviews of Modern Physics, 76, 1143,
  \dodoi{10.1103/RevModPhys.76.1143}

\bibitem[{{Rastinejad} {et~al.}(2022){Rastinejad}, {Gompertz}, {Levan}, {Fong},
  {Nicholl}, {Lamb}, {Malesani}, {Nugent}, {Oates}, {Tanvir}, {de Ugarte
  Postigo}, {Kilpatrick}, {Moore}, {Metzger}, {Ravasio}, {Rossi}, {Schroeder},
  {Jencson}, {Sand}, {Smith}, {Ag{\"u}{\'\i} Fern{\'a}ndez}, {Berger},
  {Blanchard}, {Chornock}, {Cobb}, {De Pasquale}, {Fynbo}, {Izzo}, {Kann},
  {Laskar}, {Marini}, {Paterson}, {Escorial}, {Sears}, \&
  {Th{\"o}ne}}]{Rastinejad2022}
{Rastinejad}, J.~C., {Gompertz}, B.~P., {Levan}, A.~J., {et~al.} 2022, \nat,
  612, 223, \dodoi{10.1038/s41586-022-05390-w}

\bibitem[{{Ray} {et~al.}(2022){Ray}, {Lazzati}, \& {Perna}}]{Ray2023}
{Ray}, M., {Lazzati}, D., \& {Perna}, R. 2022, arXiv e-prints,
  arXiv:2210.17507, \dodoi{10.48550/arXiv.2210.17507}

\bibitem[{{Ruiz} {et~al.}(2016){Ruiz}, {Lang}, {Paschalidis}, \&
  {Shapiro}}]{Ruiz2016}
{Ruiz}, M., {Lang}, R.~N., {Paschalidis}, V., \& {Shapiro}, S.~L. 2016, \apjl,
  824, L6, \dodoi{10.3847/2041-8205/824/1/L6}

\bibitem[{{Salafia} {et~al.}(2020){Salafia}, {Barbieri}, {Ascenzi}, \&
  {Toffano}}]{Salafia2020}
{Salafia}, O.~S., {Barbieri}, C., {Ascenzi}, S., \& {Toffano}, M. 2020, \aap,
  636, A105, \dodoi{10.1051/0004-6361/201936335}

\bibitem[{{Samsing} {et~al.}(2022){Samsing}, {Bartos}, {D'Orazio}, {Haiman},
  {Kocsis}, {Leigh}, {Liu}, {Pessah}, \& {Tagawa}}]{Samsing2022}
{Samsing}, J., {Bartos}, I., {D'Orazio}, D.~J., {et~al.} 2022, \nat, 603, 237,
  \dodoi{10.1038/s41586-021-04333-1}

\bibitem[{{Sirko} \& {Goodman}(2003)}]{Sirko2003}
{Sirko}, E., \& {Goodman}, J. 2003, \mnras, 341, 501,
  \dodoi{10.1046/j.1365-8711.2003.06431.x}

\bibitem[{{Tagawa} {et~al.}(2020){Tagawa}, {Haiman}, \& {Kocsis}}]{Tagawa2020}
{Tagawa}, H., {Haiman}, Z., \& {Kocsis}, B. 2020, \apj, 898, 25,
  \dodoi{10.3847/1538-4357/ab9b8c}

\bibitem[{{Tagawa} {et~al.}(2022){Tagawa}, {Kimura}, {Haiman}, {Perna},
  {Tanaka}, \& {Bartos}}]{Tagawa2022}
{Tagawa}, H., {Kimura}, S.~S., {Haiman}, Z., {et~al.} 2022, \apj, 927, 41,
  \dodoi{10.3847/1538-4357/ac45f8}

\bibitem[{{Thompson} {et~al.}(2005){Thompson}, {Quataert}, \&
  {Murray}}]{Thompson2005}
{Thompson}, T.~A., {Quataert}, E., \& {Murray}, N. 2005, \apj, 630, 167,
  \dodoi{10.1086/431923}

\bibitem[{{Troja} {et~al.}(2022){Troja}, {Fryer}, {O'Connor}, {Ryan},
  {Dichiara}, {Kumar}, {Ito}, {Gupta}, {Wollaeger}, {Norris}, {Kawai},
  {Butler}, {Aryan}, {Misra}, {Hosokawa}, {Murata}, {Niwano}, {Pandey},
  {Kutyrev}, {van Eerten}, {Chase}, {Hu}, {Caballero-Garcia}, \&
  {Castro-Tirado}}]{Troja2022}
{Troja}, E., {Fryer}, C.~L., {O'Connor}, B., {et~al.} 2022, \nat, 612, 228,
  \dodoi{10.1038/s41586-022-05327-3}

\bibitem[{{Ueda} {et~al.}(2003){Ueda}, {Akiyama}, {Ohta}, \&
  {Miyaji}}]{Ueda2003}
{Ueda}, Y., {Akiyama}, M., {Ohta}, K., \& {Miyaji}, T. 2003, \apj, 598, 886,
  \dodoi{10.1086/378940}

\bibitem[{{Wang} {et~al.}(2022){Wang}, {Lazzati}, \& {Perna}}]{Wang2022}
{Wang}, Y.-H., {Lazzati}, D., \& {Perna}, R. 2022, \mnras,
  \dodoi{10.1093/mnras/stac1968}

\bibitem[{{Wang} {et~al.}(2021){Wang}, {McKernan}, {Ford}, {Perna}, {Leigh}, \&
  {Low}}]{Wang2021a}
{Wang}, Y.-H., {McKernan}, B., {Ford}, S., {et~al.} 2021, \apjl, 923, L23,
  \dodoi{10.3847/2041-8213/ac400a}

\bibitem[{{Yang} {et~al.}(2022){Yang}, {Ai}, {Zhang}, {Zhang}, {Liu}, {Wang},
  {Yang}, {Yin}, {Li}, \& {L{\"u}}}]{Yang2022}
{Yang}, J., {Ai}, S., {Zhang}, B.-B., {et~al.} 2022, \nat, 612, 232,
  \dodoi{10.1038/s41586-022-05403-8}

\bibitem[{{Yang} {et~al.}(2020){Yang}, {Gayathri}, {Bartos}, {Haiman},
  {Safarzadeh}, \& {Tagawa}}]{Yang2020}
{Yang}, Y., {Gayathri}, V., {Bartos}, I., {et~al.} 2020, \apjl, 901, L34,
  \dodoi{10.3847/2041-8213/abb940}

\bibitem[{{Zhu} {et~al.}(2021{\natexlab{a}}){Zhu}, {Wang}, \&
  {Zhang}}]{Zhu2021nu}
{Zhu}, J.-P., {Wang}, K., \& {Zhang}, B. 2021{\natexlab{a}}, \apjl, 917, L28,
  \dodoi{10.3847/2041-8213/ac1a17}

\bibitem[{{Zhu} {et~al.}(2021{\natexlab{b}}){Zhu}, {Zhang}, {Yu}, \&
  {Gao}}]{Zhu2021a}
{Zhu}, J.-P., {Zhang}, B., {Yu}, Y.-W., \& {Gao}, H. 2021{\natexlab{b}}, \apjl,
  906, L11, \dodoi{10.3847/2041-8213/abd412}

\end{thebibliography}

\end{document}